\documentclass[12pt,twoside,english]{article}
\usepackage{ifthen}                % if then else Abfragen
\newboolean{boldeqnitalic}
\setboolean{boldeqnitalic}{true}

\usepackage{caption}
\usepackage[intlimits] {amsmath}   % muss for defjs# stehen 
\usepackage{defjs2}                % eigene Definitionen    
\usepackage{epsfig}
\usepackage{psfig}
\usepackage{graphicx}
\usepackage{psfrag}
\usepackage[usenames]{color}
\usepackage{colortbl}
\usepackage{ifthen}
\usepackage{fancyhdr}
\usepackage{rotating}
\usepackage{multirow}
\usepackage{booktabs}              % Dicke Tabellenlinien
\usepackage{amssymb}
\usepackage{amsbsy}
\usepackage{bm}                    % fette Schrift in Formeln
\usepackage{babel}
\usepackage{theorem}
\usepackage{upgreek}  % gerade (roman) griechische Buchstaben z.B. \upalpha
\usepackage{pifont}   % zusaetzliche Schriften, z.B. eingekreiste Zahlen
\usepackage{hyperref}
\usepackage{mathrsfs}
\usepackage{datetime}
\usepackage{tikz}
\usetikzlibrary{matrix}
\usepackage[all]{xy}
\usepackage{rotating}
\usepackage{subfigure}
\definecolor{dunkelgrau}{rgb}{0.8,0.8,0.8}
\definecolor{hellgrau}{rgb}{0.90,0.90,0.90} %... slightly darker 
\usepackage{natbib} 
\fancypagestyle{plain}{%
  \fancyhead{}%
  \fancyfoot[c]{\sffamily\thepage}%
}
\makeatletter                           % Leere Fuellseiten
\def\cleardoublepage{\clearpage\if@twoside \ifodd\c@page\else
  \hbox{}
  \vspace*{\fill}
  \thispagestyle{empty}
  \newpage
  \if@twocolumn\hbox{}\newpage\fi\fi\fi}
\makeatother
\begin{document}
\unitlength1.0cm
\frenchspacing

\thispagestyle{empty}
 
\vspace{-3mm}
\begin{center}
  {\bf \large Tibial Implant Fixation in TKA Worth A Revision?}  \\[2mm]
%  {\bf \large Standard Tibial Plate-Stem Component Design in}  \\[2mm]
%  {\bf \large Total Knee Replacement Worth a Revision?} \\[2mm]
%  {\bf \large A Finite Element Analysis Following a Bionics Perspective} \\[2mm]
%{\bf \large Towards Physiological Conditions by a Bionics Guideline} \\[2mm]
{\bf \large -- How to Avoid Stress-Shielding } \\[2mm]
{\bf \large Even for Stiff Metallic Implants} \\[2mm]
\end{center}

\vspace{4mm}
\ce{B. Eidel$^{\,a,\ast}$, A. Gote$^{\,a}$, C.-P. Fritzen$^{\,b}$, A. Ohrndorf$^{\,c}$, H.-J. Christ$^{\,c}$}

\vspace{4mm}

\ce{\small $a$ \, Heisenberg-Group, Institut f\"ur Mechanik} 
\ce{\small University Siegen, 57068 Siegen, Paul-Bonatz-Str. 9-11, Germany} 
\ce{\small $^{\ast}$e-mail: bernhard.eidel@uni-siegen.de
%, phone: +49 271 740 2224, fax: +49 271 740 2436
} 
\vspace{2mm}
\ce{\small $b$ \, Institut f\"ur Mechanik}
\ce{\small Universit\"at Siegen, 57068 Siegen, Paul-Bonatz-Str. 9-11, Germany}
\vspace{2mm}
\ce{\small $c$ \, Institut f\"ur Werkstofftechnik}
\ce{\small Universit\"at Siegen, 57068 Siegen, Paul-Bonatz-Str. 9-11, Germany}

%\ce{\small Department Mechanical Engineering, Faculty IV} 
%\ce{\small Universit\"at Siegen, 57068 Siegen, Paul-Bonatz-Str. 9-11, Germany}

\bigskip

\begin{center}
{\bf \large Highlights}

\bigskip

{\footnotesize
\begin{minipage}{15.5cm} 

\vspace*{-2mm}

\begin{itemize}
 \item Force decomposition into plate and stem parts is an indicator for post-surgery tibial bone loss.    
 \\[-6mm] 
 \item Force transfer mainly through the tibial plate (stem) reduces (increases) stress-shielding. \\[-6mm]
 \item Strong activation of the plate in force transfer by sliding friction at the stem-bone interface.  
 \\[-6mm]
 \item Stress-shielding can be overcome even for stiff metallic implants.  
\end{itemize}

\end{minipage}
}
\end{center}

%\bigskip
  
\begin{center}
{\bf \large Abstract}

\bigskip

{\footnotesize
\begin{minipage}{14.5cm} 
\noindent
In total knee arthroplasty (TKA) force is transmitted into the tibia by a combined plate-stem device along with cemented or cementless stem fixation. The present work analyzes this force transmission in finite element simulations with the main aim to avoid reported postsurgical bone density reduction as a consequence of a reduced tibial bone loading. In the numerical analysis different implant materials, stem/extension lengths and   implant-to-stem interface conditions are considered, from a stiff fully cemented fixation to sliding contact conditions with a low friction coefficient. The impact of these variations on bone loading changes are measured by (i) decomposing the total force into parts mediated by the plate and by the stem and by (ii) post-surgery strain energy density (SED) deviations. Based on a bionics-inspired perspective on how nature in pre-operative conditions carries out force transfer from the knee joint into the tibia, a modified implant-bone interface is suggested that alters force transmission towards physiological conditions while preserving the geometries of the standard plate-stem endoprosthesis design. The key aspect is that the axial force is predominantly transmitted through the plate into proximal bone which requires a compliant bone-stem interface as realized by sliding friction conditions at a low friction coefficient. These interface conditions avoid stress shielding almost completely, preserve pre-surgery bone loading such that bone resorption is not likely to occur.
\end{minipage}
}
\end{center}

{\bf Keywords:}
Total knee arthroplasty; Stress shielding; Tibia; Implant-bone interface; Finite element analysis \hfill %vers.\,\today \, at \currenttime\\
%\hspace*{6.3cm} 
%\\[2mm]
%\hfill vers.\,\today \, at \currenttime\\
\begin{center}
\textit{26.09.2020 accepted for publication in} Computer Methods in Biomechanics and Biomedical Engineering\\
\textit{The finally published version can show minor differences to the present preprint.}
\end{center}
\vfill
\newpage 
%\input{Frontpage-2}

%----------------------------------------------------------------------------------------------------------------

%\input{introduction}

\section{Introduction}
\label{sec:intro}
%{\color{blue} The number of total knee replacements performed each year in the United States has increased dramatically, from 31.2 per 100\,000 person-years during the period 1971--1976 to 220.9 during the period 2005--2008. In 2012, more than 670,000 total knee replace-ments  were  performed  in  the  United  States  alone, with corresponding aggregate charges of \$36.1 billion. from: Skou et al.: A Randomized, Controlled Trial of Total Knee Replacement. The New England Journal of Medicine vol. 373 no. 17, 2015}

Total knee arthroplasty (TKA) has evolved into a mature surgical intervention  {\color{black} and the number of patients undergoing TKA has considerably increased in the last decades. In the United States for example, the number of TKA performed each year has increased from 31.2 per 100\,000 person-years during the period 1971--1976 to 220.9 during the period 2005--2008. In 2012, more than 670,000 total knee replacements were  performed in the United States alone, with corresponding aggregate charges of \$36.1 billion \cite{Skou.2015}.} 
%Skou et al.: A Randomized, Controlled Trial of Total Knee Replacement. The New England Journal of Medicine vol. 373 no. 17, 2015.}

Despite long survivorship (survival rates greater than 90\% after 10 years follow-up), the increasing number of primary TKA has been associated with increased rates of revision TKA procedures. According to \cite{Sharkey.2014} the most common failure mechanisms {\color{black} among the revision cases} of TKA was loosening (39.9\%). 

In view of these numbers, most notably the percentage of loosening among the failure mechanisms, there is a need and space for improvements. While implant loosening can have different causes as implant wear or infection \cite{Carr.2012}, stress shielding inducing bone loss an important case likewise, for a review of periprosthetic osteolysis we refer to \cite{Gallo.2013}. 

The overall aim of this paper is to overcome stress shielding and bone resorption in post-TKA tibial bone by a novel implant fixation concept. To put things into perspective and to underpin the novelty of the present approach in comparison to other works, previous and on-going research activities shall be classified and their results briefly summarized.
\\[2mm] 
Candidates to cause tibial stress shielding and bone resorption in TKA are (i) the implant material, (ii) the stem length, (iii) the fixation concept -- surface cementation versus full cementation, (iv) the baseplate positioning, and more characteristic features of TKA; for their investigation results from clinical research and through simulations have been obtained, partially with controversial results and conclusions. 
\\[2mm] 
One controversial topic is the role of implant material. 
\cite{Martin.2017} considered medial tibial stress shielding by a radiographic comparative analysis with a focus on the role of material stiffness, cobalt chromium (CoCr), or an all polyethylene (AP) tibial implant. The finding that the CoCr cohort had the highest amount of medial tibial bone loss, the AP cohort the least amount of stress shielding, suggests that the more compliant the implant, the less the bone loss. 

In their finite element analysis (FEA) \cite{Au.2007} found that the Young's modulus mismatch of implant compared to bone is not sufficiently descriptive as the primary cause of stress shielding. Instead, they pointed out that loading conditions as a result of altered bone or implant condylar surface geometry, load placement on the condylar surface, and load pattern created by the TKA are at least as important in observed stress shielding.

The results of \cite{Zhang.2016} suggest that the effect of stress shielding on the periprosthetic bone is more significantly influenced by the implant material than by the implant geometry. Pronounced stress shielding was found in metal-backed (MB) cases as opposed to AP cases, particularly in the bone regions right beneath the baseplate. 
\\[2mm]
The influence of stem design and stem length must not be restricted to stress shielding but must cover the aspect of mechanical stability likewise. \cite{Completo.2009} assessed stress shielding and stress concentrations through FEA of different tibial stem designs. For bone regions underneath the tibial tray and around the stem, the geometry of stem had a more pronounced bone effect compared to the stem material. The results of this study support that short stems produce a minor effect in bone relatively to long stem in terms of stress shielding and stress concentration at tip region. No significant stress shielding differences was observed between Co–Cr and titanium stems. Overall, all stems provoked high stress concentrations in bone at the tip of the stem.

\cite{Scott.2012} analyzed the role of stems in TKA. They found that stems improve the mechanical stability of tibial components (resistance to shear, reduced tibial lift-off, reduced micro motion), but come at a cost of stress shielding along their length. The authors conceded the disadvantages including stress shielding along the length of the stem with associated reduction in bone density and a theoretical risk of subsidence and loosening, periprosthetic fracture and end-of-stem pain. While these features make long stems unattractive in the primary TKA setting, they are often desirable in revision surgery with bone loss and instability.  
\\[2mm] 
\cite{Innocenti.2009} investigated the influence of baseplate positioning; they highlighted the medial cortical support of the tibial baseplate for normal stress transfer to the underlying bone which prevents local bone resorption at the proximal tibia as a result of  stress shielding.
\\[2mm] 
Another important aspect of stress-shielding in TKA is cementation in its variants of full versus surface cementation, and other parameters like layer thickness and bone-cement interface. \cite{Cawley.2013} found that 
%higher cortical strains are measured for surface cementation, as full cementation creates a stiffer proximal tibial structure. Simulations reveal that both cementation techniques result in lower cancellous stresses under the baseplate compared to the intact tibia, with greater reductions being computed for full cementation. The surface cementation model displays the closest cancellous stress distribution to the intact model. In addition, bone remodeling simulations predict more extensive bone resorption under the baseplate for full cementation (43\%) than for surface cementation (29\%). Interpretation: 
full cementation results in greater stress reduction under the tibial baseplate than surface cementation, and concluded that surface cementation will result in less proximal bone resorption, thus reducing the possibility of aseptic loosening. 

\cite{Gallo.2013} investigated the role of the bone-cement interface,  \cite{Vanlommel.2011,Cawley.2012,Cawley.2013} address the question of the best technique for cementation in TKA and thereby point to inter-surgeon differences which might introduce a bias thus avoiding a consensus in that question. \cite{Schlegel.2015} highlight the crucial role of surface preparation and pulsed lavage in TKA.   
\\[2mm] 
With the same aim as the above cited works, namely to overcome bone resorption in post-TKA tibial bone, the present work suggests a new pathway to achieve that goal by proposing a novel implant fixation concept.

%\vfill
%\newpage

%======================================================================================================
\section{Materials and Methods}
\label{sec:Methods}
%======================================================================================================

\subsection{Tibia reconstruction}
\label{subsec:FemurReconstruction}
 
The CAD-model of the tibia was generated using trial version of \textit{Simpleware Software} developed by \textit{Synopsys Inc., USA}. The CT data used for the present study were obtained from \cite{VHP-ctdata}, \cite{VHP-general}, and belong to a 59 year old female cadaver. A data set of 340 CT images ($512\times512$ pixels, 12 bit gray scales, DICOM format) was used for the tibia reconstruction. The vertical and horizontal spacing between the pixels each of 0.33\,mm enabled cubic voxels in reconstruction. 

The heterogeneous Young's modulus distribution is obtained from the computer tomography (CT) images in three steps. Firstly, the CT number measured in terms of the so-called Hounsfield unit (HU) is calculated for bony tissue by means of the attenuation coefficients $\mu$ of both bony tissue and water; it holds 

\begin{equation}
CT(\mu_\textrm{tissue}):=\frac{\mu_\textrm{tissue}-\mu_\textrm{water}}{\mu_\textrm{water}} \cdot 1000\,\textrm{HU}\,,
\label{CTnorm}
\end{equation}

Next, the tissue density $\rho_\textrm{tissue}$ is calculated as a function of the $CT$ number

\begin{equation}
\rho_\textrm{tissue}(CT)=\frac{\rho_\textrm{water}-\rho_\textrm{air}}{CT_\textrm{water}-CT_\textrm{air}}\cdot CT_\textrm{tissue}+\rho_\textrm{water}\,.
\label{dichte1}
\end{equation}

The linear conversion in \eqref{dichte1} reflects the fact that the $CT$ number is linear in the density of the material; $CT$ values for water and air are known.

Using the conversion of \cite{Rho.1995} for the proximal tibia, the density calculation in \eqref{dichte1} obeys the form \eqref{CTtoRho} and the Young's modulus is calculated according to \eqref{RhotoE}. The conversions are carried out using Bonemat tool developed by \cite{Taddei.2007}   
\begin{eqnarray}
\rho_\textrm{tissue}(CT) &=& 114 + 0.916 \cdot CT \, \, \text{(kg/m$^{3}$)} \, .
\label{CTtoRho} \\
E &=& -326+5.54 \cdot \rho \, \, \text{(MPa)}
\label{RhotoE}
\end{eqnarray}

The data sets for the reconstructed bone model and finite element discretizations are published as a supplement. It shall facilitate research in the field of knee implants in the spirit of \cite{Viceconti.1996} and \cite{Eidel.2018}.

%------------------------------------------------------------------------------------------------------
\subsection{Finite element analysis}
\label{subsec:FEA} 
%------------------------------------------------------------------------------------------------------
The finite element solver Abaqus 2017 (Dassault Syst\`{e}mes, Paris, France) was used in geometrically nonlinear simulations for deformation and stress analyses. 
Tetrahedral finite elements with quadratic shape functions (10 nodes) along with displacement degrees of freedom were used for the discretization of bone and implant. 

\subsubsection{Boundary and interface conditions} 

\begin{Figure}[htbp]
	\begin{minipage}{16.5cm}  
		\centering
		\subfigure[]{\includegraphics[height=7cm, angle=0]{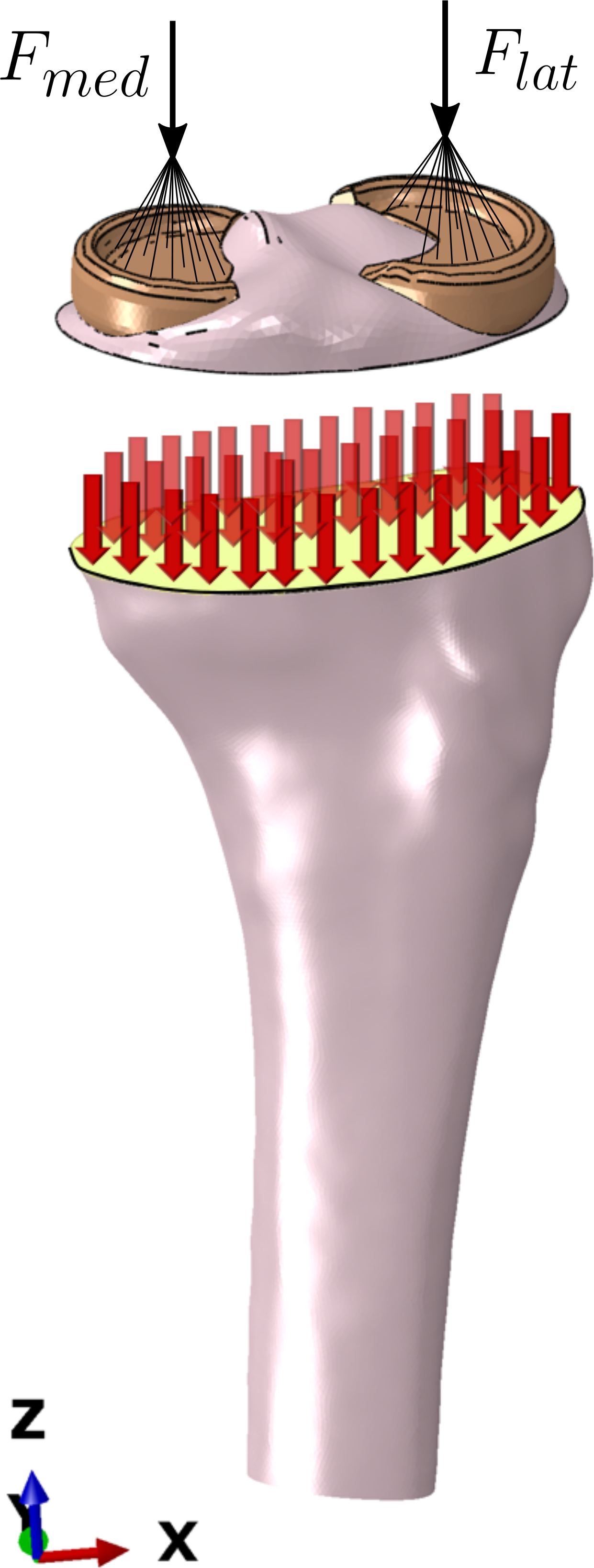}} \hspace*{6mm}
		\subfigure[]{\includegraphics[height=7cm, angle=0]{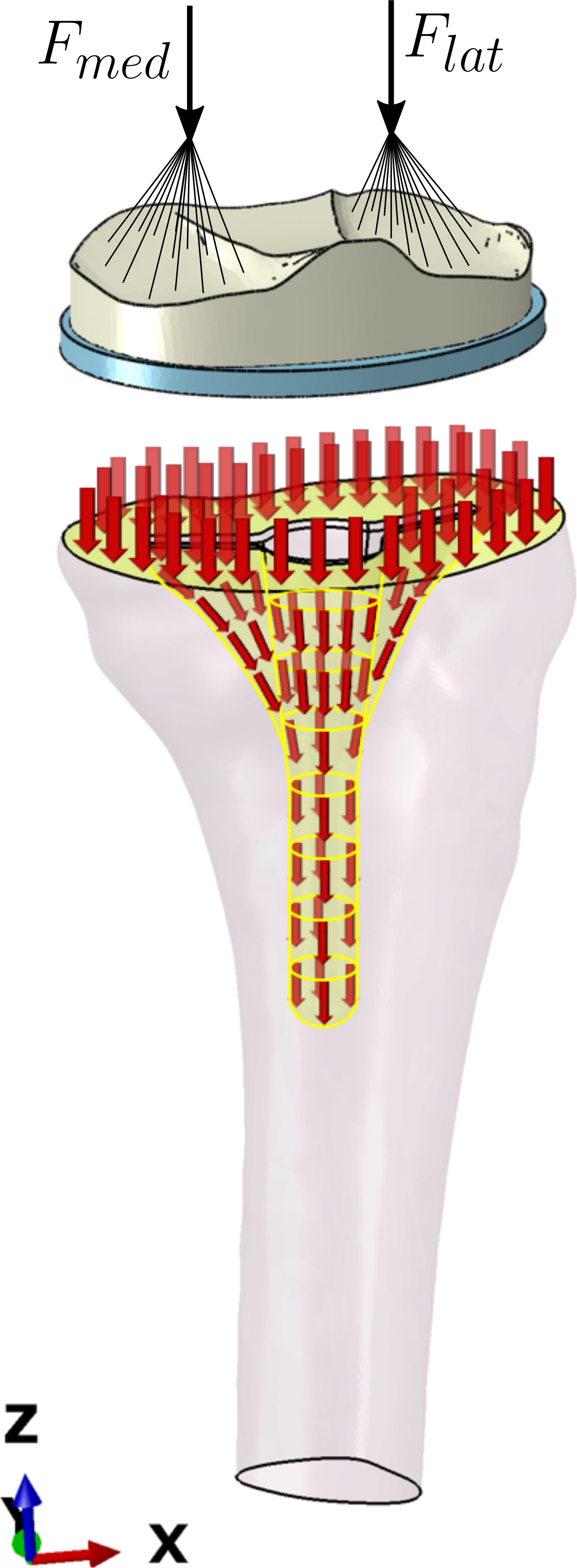}} \hspace*{6mm}
		%\subfigure[]{\includegraphics[height=7cm, angle=0]{Pictures/Cemented_Tibia.png}} \hspace*{4mm}
		\subfigure[]{\includegraphics[height=6cm, angle=0]{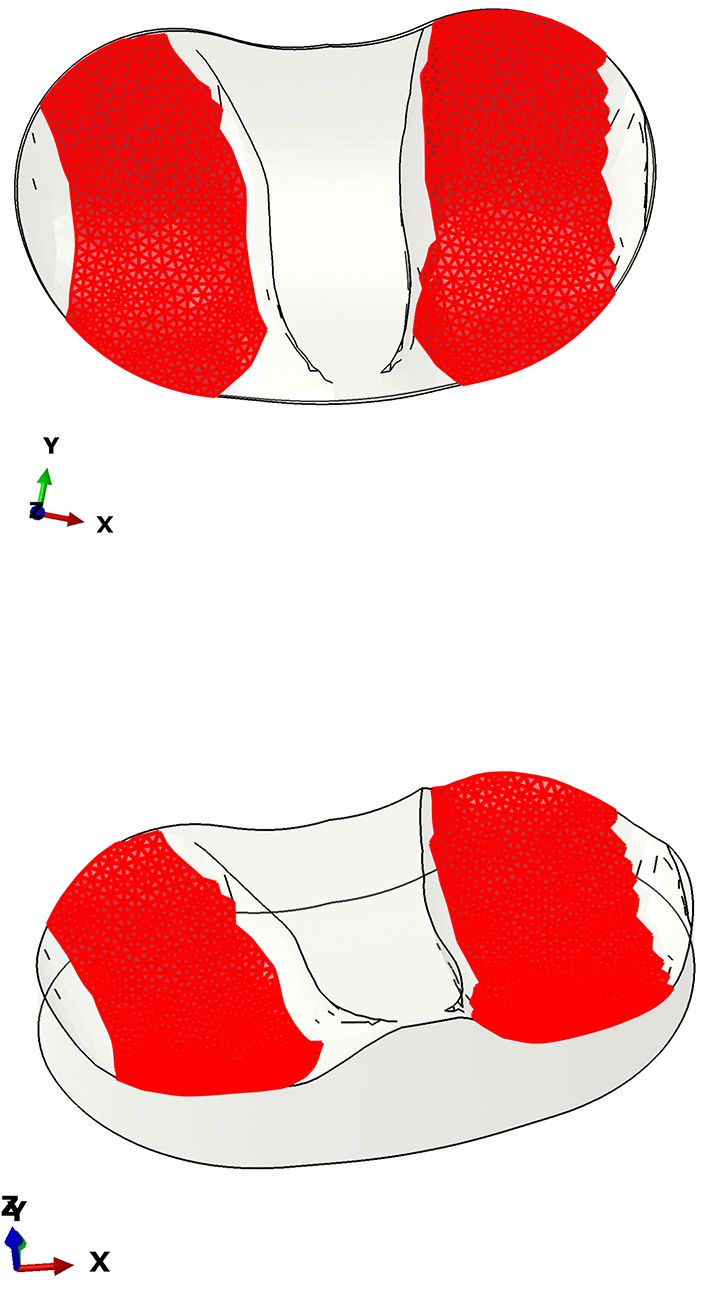}}  
	\end{minipage}
	\caption{(a) Pre-surgery tibial BCs, full force transmission in the cross-section. (b) Post surgery tibial BCs, force transmission decomposed into plate part and stem part.(c) Medial and lateral surfaces of PE insert element on which the respective loads are acting, contact areas in red color. {\color{black}Notice, that --opposed to the sketches (a) and (b) the real stress distributions in a cutting plane of the proximal tibia is strongly nonuniform due the heterogeneity of stiffness, compare Fig.  \ref{tab:YoungsMod-SEDpresurg-Tibia}. The contact area marked in (c) describes the maximal range for a variety of movements; the true contact area in a particular position is considerably smaller.} \label{fig:BVP_Tibia_and_concept}}
\end{Figure}

Figure~\ref{fig:BVP_Tibia_and_concept} {\color{black}schematically} displays the difference in the load-transfer of pre-surgery to post-surgery axial load transfer. Under pre-surgery conditions the joint forces applied by the two femoral condyles are completely transmitted by stresses in the cross-sectional plane which in surgery is chosen as the tibial resection plane. Under post-surgery conditions the axial force is decomposed according to \eqref{eq:Plate-stem-force-decomposition}, partially transmitted by the plate through normal stresses, partially by the stem through shear stresses  
\begin{equation}
\bm F_{\mbox{\scriptsize axial}} = \bm F_{\mbox{\scriptsize axial,plate}} + 
\bm F_{\mbox{\scriptsize axial,stem}} \, .
\label{eq:Plate-stem-force-decomposition}
\end{equation}

The force decomposition between plate and stem parts generally depends on stiffness of bone and implant, interface material (cemented or cement free fixation), and stem extension length. It is however clear that the force transmitted at the cross section of the base plate is in either case lowered compared to pre-surgery conditions, which implies a loss of bone loading in the proximal regions. 

It is the main aim of the present work (i) to quantify the force decomposition according to \eqref{eq:Plate-stem-force-decomposition} for the standard fixation concept, (ii) to analyze its consecutive force flow through the tibia along with a comparison to pre-surgery bone loading, and (iii) to propose a novel fixation concept that minimizes post-TKA loading reduction. 

Dirichlet boundary conditions are chosen at the distal end of the modeled tibia that is all $D=3$ degrees of freedoms are fixed. For the Neumann conditions, only the forces acting at the tibia plate in $z$-direction are modeled with $|\bm F|=F_z= 2 \times 543$ N, {\color{black}\cite{Bergmann.2014}, \cite{Bergmann.2020}}. The set-up of knee joint prostheses that enable in vivo measurements of loading are described in \cite{Damm.2010}. 

For the bone-implant interfaces we analyze various boundary conditions as detailed in Sec.~\ref{subsubsec:Standard-Short-Stem-Hip-Endoprothesis}. In either case, sticking friction conditions are set for the interface between the tibial tray and bone. For the case of a cemented fixation we assume sticking friction for the interfaces of implant-cement and cement-bone. For the cementless fixation we consider the cases of sticking friction and two different coefficients of friction.

\subsubsection{Material properties} 
%------------------------------------------------------------------------------------------------------

For the material behavior of tibial bone, an isotropic linear elasticity law is assumed to hold. The assumption of linear elasticity is corroborated by recent experimental findings by \cite{Juszczyk.2011} and \cite{Grassi.2016} (for femoral bone), the validity of isotropic elasticity in simulations is underpinned by \cite{Schileo.2014}. {\color{black} Cowin presents orthotropic elasticity parameters for the tibia, which are however throughout constant and thus do not account for the pronounced non-homogeneity of real bone \cite{Cowin.2009}.} The figures of the strongly heterogeneous Young's modulus distribution follow from bone reconstruction as described in Sec.~\ref{subsec:FemurReconstruction}. The Poisson's ratio is assumed to be constant, $\nu=0.3$. For the implant material, linear elastic, isotropic material parameters are chosen for both the titanium-alloy (TiAl6V4) \cite{Long.1998} as well as for the cobalt-chrome alloy (CoCr) version. 
The very standard of bone cement is PMMA (Polymethyl methacrylate). Polyethylene (PE) is used for the tibial tray; moreover, we consider for its low stiffness a fictitious implant fully made of PE (all-PE). The material parameters of linear elasticity are listed in Table~\ref{tab:Material-parameters}.

\begin{Table}[htbp]
	\begin{minipage}{16.5cm}  
		\centering
		\renewcommand{\arraystretch}{1.2} 
		\begin{tabular}{lccccc}
			\hline
			&  Ti &  CoCr & PE & PMMA & Bone \\ 
			\hline
			Young's modulus $E$ [MPa] & {110\,000}  & {210\,000}  &    {1\,200} & 2\,200 & { $\in$ [30,\, 8\,193.59] }  \\
			Poisson's ratio $\nu$     & {0.33}    & {0.36}        &    {0.46} & 0.3  & {0.3} \\
			\hline
		\end{tabular}
		\newline 
	\end{minipage}
	\caption{Elasticity constants for the tibial implant components (plate and stem), for the cement layer (PMMA), and for tibial bone. \label{tab:Material-parameters}}
\end{Table}

\subsubsection{Variants of tibia implant system}
\label{subsubsec:Standard-Short-Stem-Hip-Endoprothesis} 
%------------------------------------------------------------------------------------------------------ 
In the numerical analysis we consider distinct cases of the tibial implant systems showing variations in geometry, material and fixation:
\begin{enumerate}
	\item Stem extension length: $l=$\{5,40,75\} mm, Fig.~\ref{fig:Variants-of-endoprotheses} (b,c,d). {\color{black}Stem thickness: 10 mm.} 
	\item Material: titanium alloy (Ti), cobalt-chrome alloy (CoCr), all-polyethylene (PE). The first two are frequently referred to as metal-backed (MB).
	\item Implant-to-bone interface conditions: variants of sticking friction conditions and sliding friction conditions are considered. For sticking friction, the variant of (i) full cementation (FullC) and (ii) surface cementation (SurfC) along with stem fixation through press-fit and, after trabecular ingrowth, through osseointegration. For the variant of sliding friction different cases of coefficient of frictions (cof) are compared, (iii) cof=0.2 and the idealized case of (iv) cof=0.0 for perfectly smooth interfaces.  
\end{enumerate}

Figure~\ref{fig:Variants-of-endoprotheses} (a) displays the orientation of the resection plane to the mechanical axis and the thickness of maximal bone removal, the image in (e) shows the case of full cementation.

\begin{Figure}[htbp]
	\begin{minipage}{16.5cm}  
 	\centering
 	\subfigure[Resection plane]{\includegraphics[height=6.5cm, angle=0]{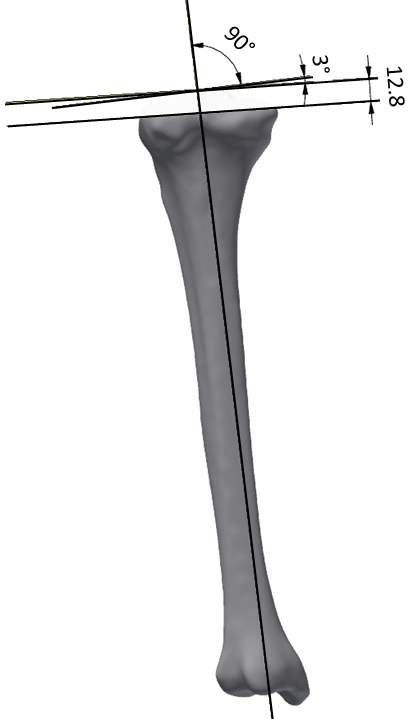}}  \hspace*{0mm}
 	\subfigure[5\,mm]{\includegraphics[height=5.3cm, angle=0]{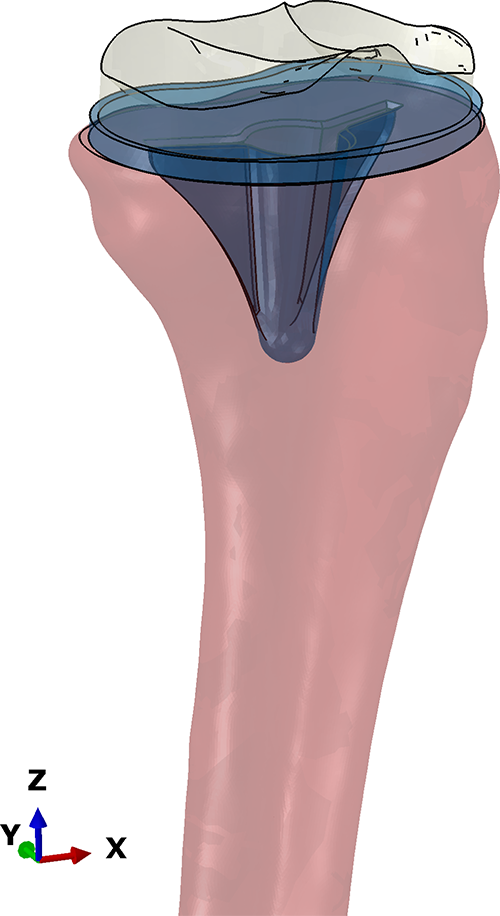}}  \hspace*{0mm}
 	\subfigure[40\,mm]{\includegraphics[height=5.3cm, angle=0]{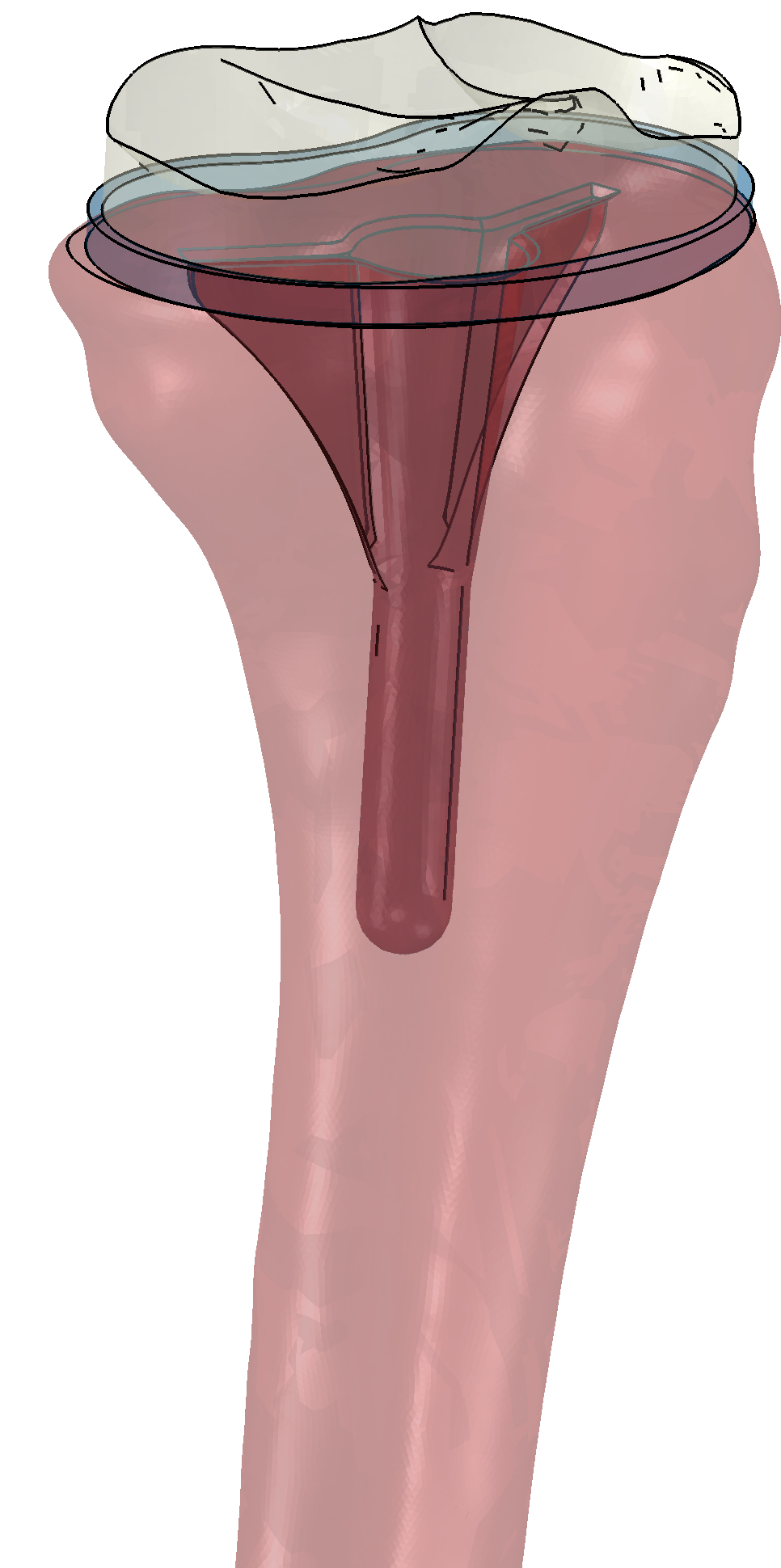}} \hspace*{0mm}
 	\subfigure[75\,mm]{\includegraphics[height=5.3cm, angle=0]{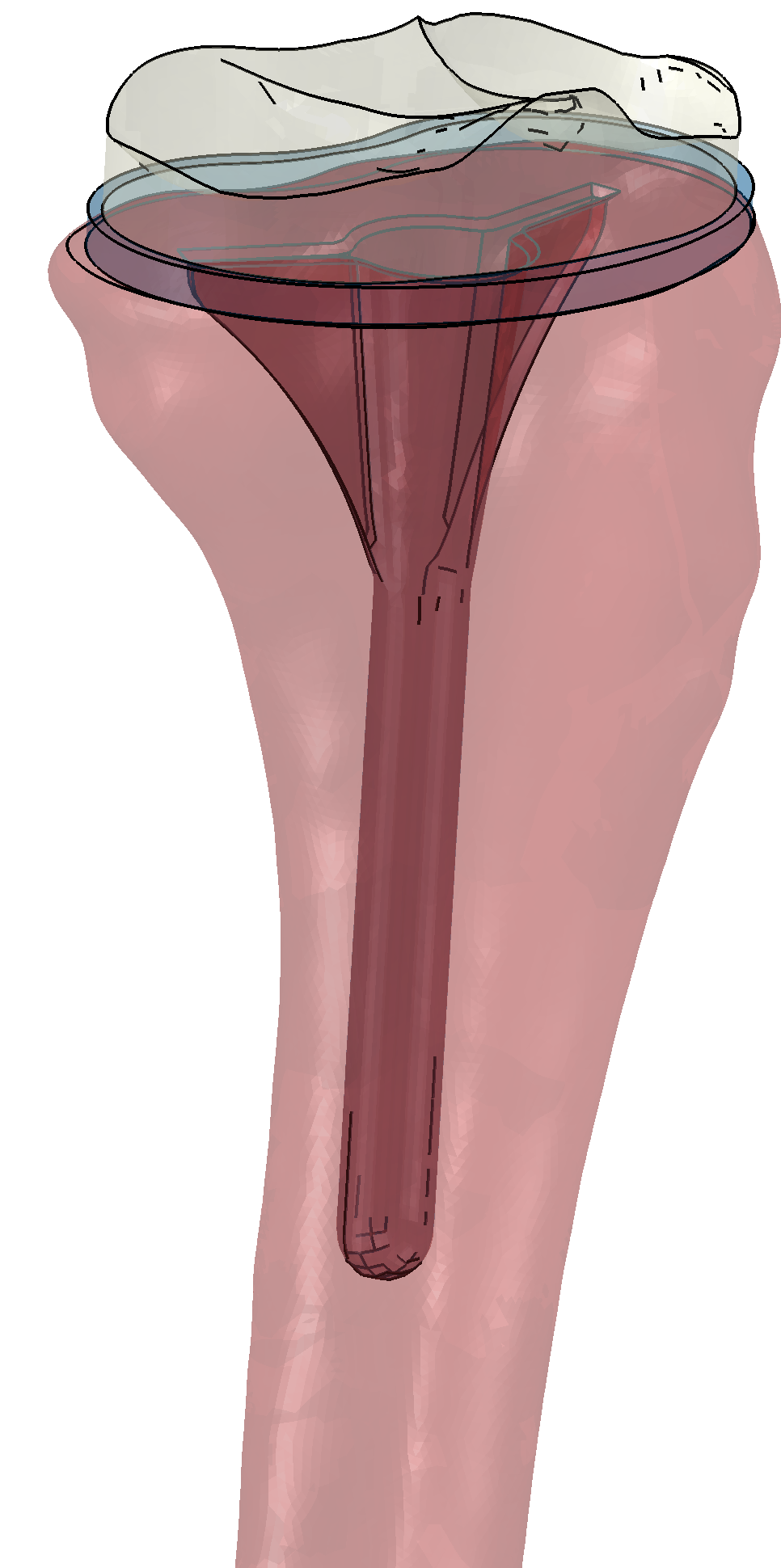}} \hspace*{1mm} 
 	\subfigure[FullC]{\includegraphics[height=5.8cm, angle=0]{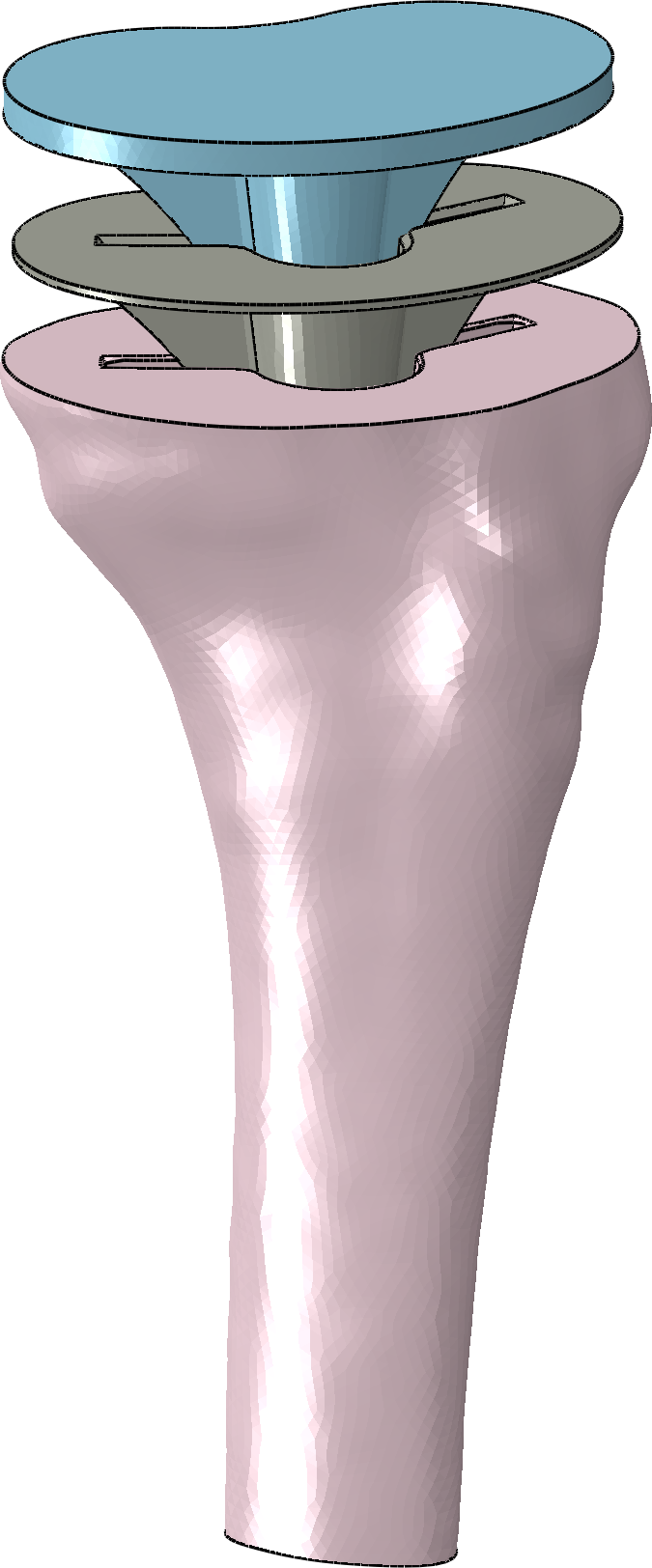}} 
		%\subfigure[]{\includegraphics[height=4.5cm, angle=0]{surface-coverage_MD_cut.eps}} 
	\end{minipage}
	\caption{{\bf Tibial device:} (a) Orientation of resection plane, (b-d) tibial device with base plate and PE insert on top of it, the two fins for rotational stiffness and conical stems with extensions of various lengths, (e) the fully cemented case.
		\label{fig:Variants-of-endoprotheses}}
\end{Figure}
 
%\begin{Table}[htbp]
%		\centering
%		\renewcommand{\arraystretch}{1.2} 
%		\begin{tabular}{lccc}
%			\hline
%			Variant            &   5mm     &  40mm    & 75mm     \\ 
%			\hline
%			No. of nodes        & {406035}  & {320916} & {813233}   \\
%			No. of elements     & {266569}  & {207313} & {559127}   \\
%			\hline
%		\end{tabular}
%	\caption{ FE discretization for different variants of post surgery tibia. \label{tab:FE-discretization}}
%\end{Table}   

\subsubsection{Measures in the numerical analysis} 
%-----------------------------------------------------------------------------------------------------------------------------------------------------------------

The strain energy density (SED), which is defined by the scalar-product of work-conjugate pairs of stress and strain tensors, here the Second Piola-Kirchhoff stress tensor $\bm S$ and the Green-Lagrange strain tensor $\bm E$,   
\begin{equation}
\mbox{SED} = \frac{1}{V}\int_V \bm S:\bm E\,\mbox{dV} 
\label{eq:SED}
\end{equation}
is widely accepted as a mechanical stimulus for bone remodeling in biomechanics \cite{Huiskes.1992, Ehrlich.2002}, 
for a detailed discussion see \cite{Turner.1998}, \cite{Ambrosi.2011}.

The relative, percental SED deviation of the post-TKA bone from the pre-TKA bone is calculated according to 
\begin{equation}
\textrm{SED}_{\mbox{\scriptsize diff}} = 
\dfrac{ \textrm{SED}^{\mbox{\scriptsize post-TKA}} - \textrm{SED}^{\mbox{\scriptsize pre-TKA}} }
{\textrm{SED}^{\mbox{\scriptsize pre-TKA}}}\,.
\label{eq:SED-difference-definition}
\end{equation}

A similar metric based on von-Mises stress instead of SED is used by \cite{Fraldi.2010} --referred to as Stress Shielding Intensity SSI-- and by \cite{Boyle.2011}.

For a comparison among the implant variants and with the pre-surgery state, SED differences according to \eqref{eq:SED-difference-definition} and a decomposition of force transfer into plate and stem parts.

%======================================================================================================================
\section{Results}
\label{sec:results} 
%======================================================================================================================
  
%------------------------------------------------------------------------------------------------------------------------   
\subsection{Presurgery load transfer}
\label{subsec:Load-transfer-preTHA} 
%-----------------------------------------------------------------------------------------------------------------------------------------------------------------

\begin{Table}[htbp]
	\begin{tabular}{| >{\centering\arraybackslash} m{8mm} | >{\centering\arraybackslash} m{6.5cm} >{\centering\arraybackslash} m{6.5cm}  |}  
		\hline 
	    & Spongiosa  & Corticalis \\ 
		\hline
		& & \\ 
		$E$ & {\includegraphics[height=9.0cm, angle=0]{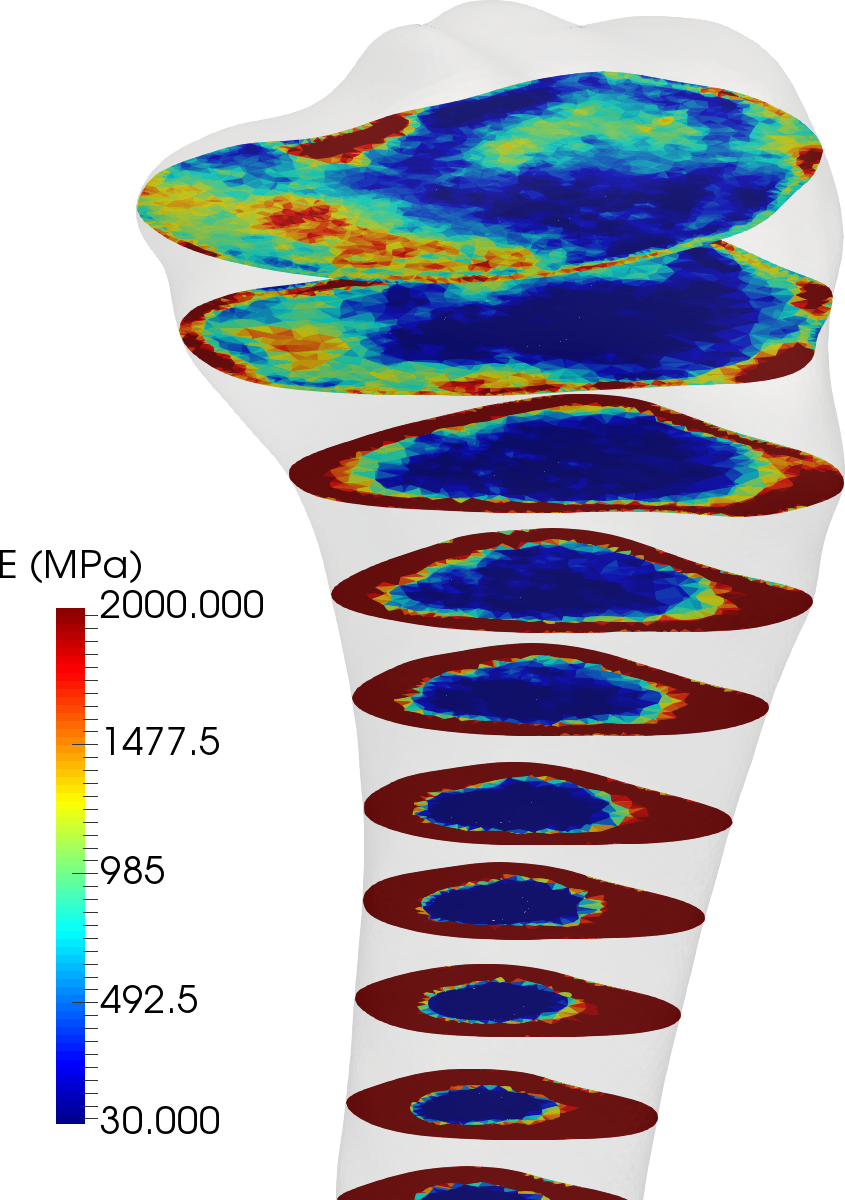}}
	  	    & {\includegraphics[height=9.0cm, angle=0]{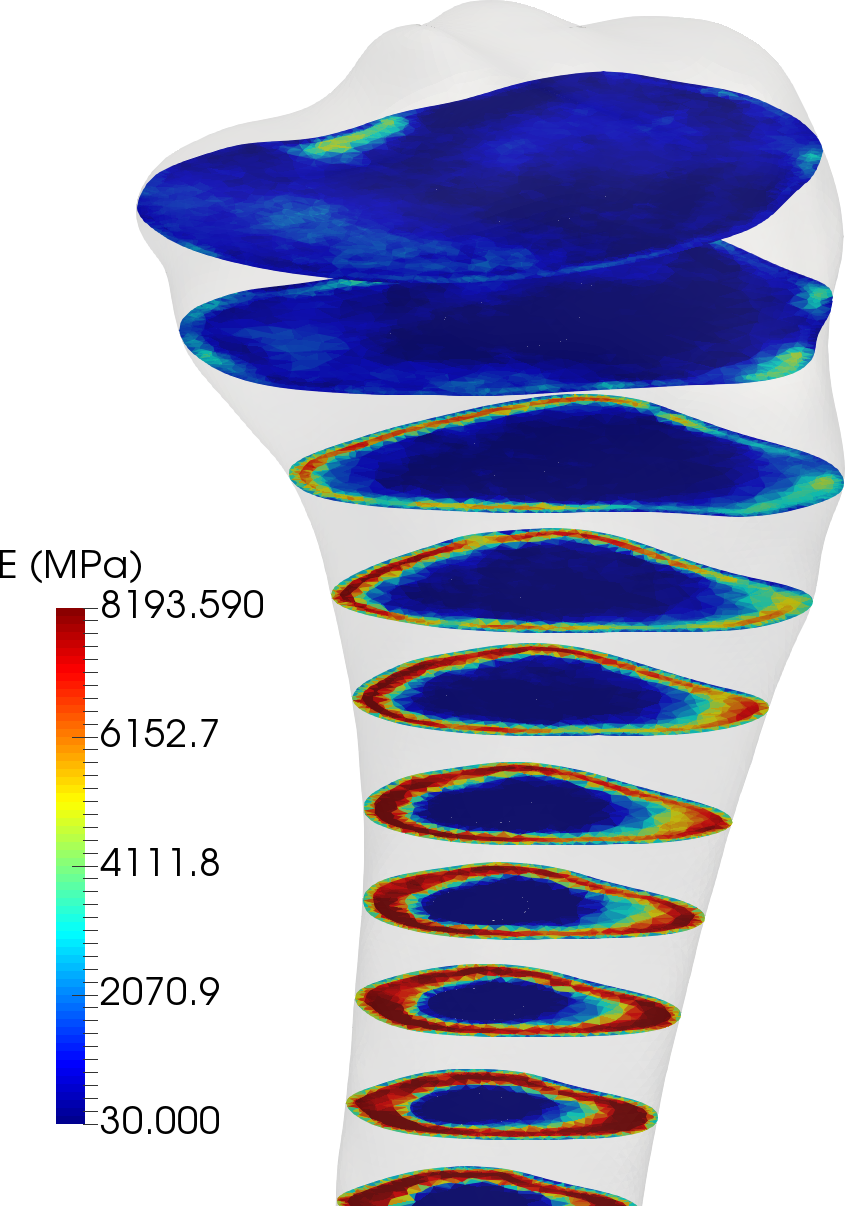}}
	    \\ 
		\hline  
		& &  \\ 
		SED  & \hspace*{4mm} {\includegraphics[height=8.4cm, angle=0]{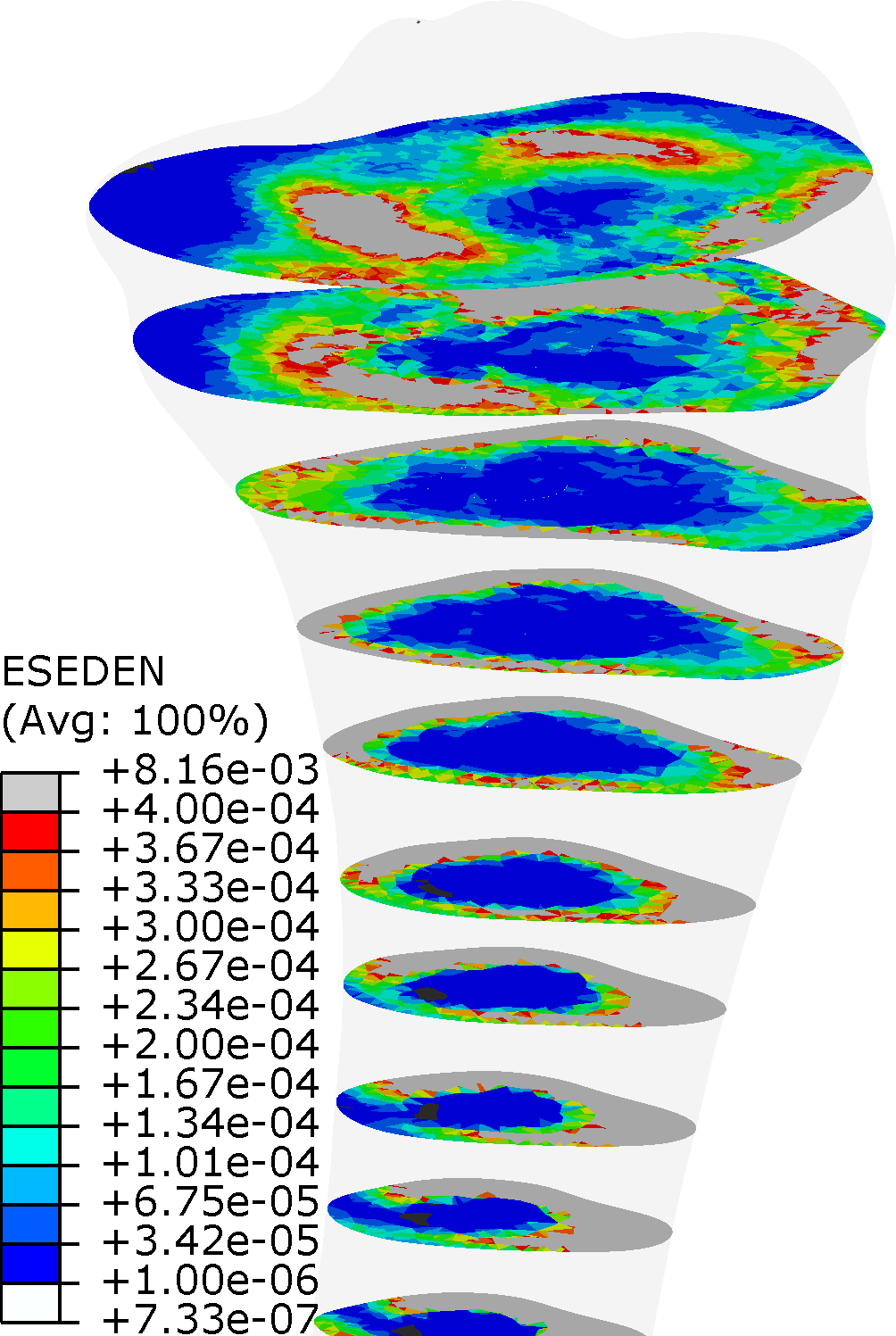}}
	      	 & \hspace*{0mm} {\includegraphics[height=8.4cm, angle=0]{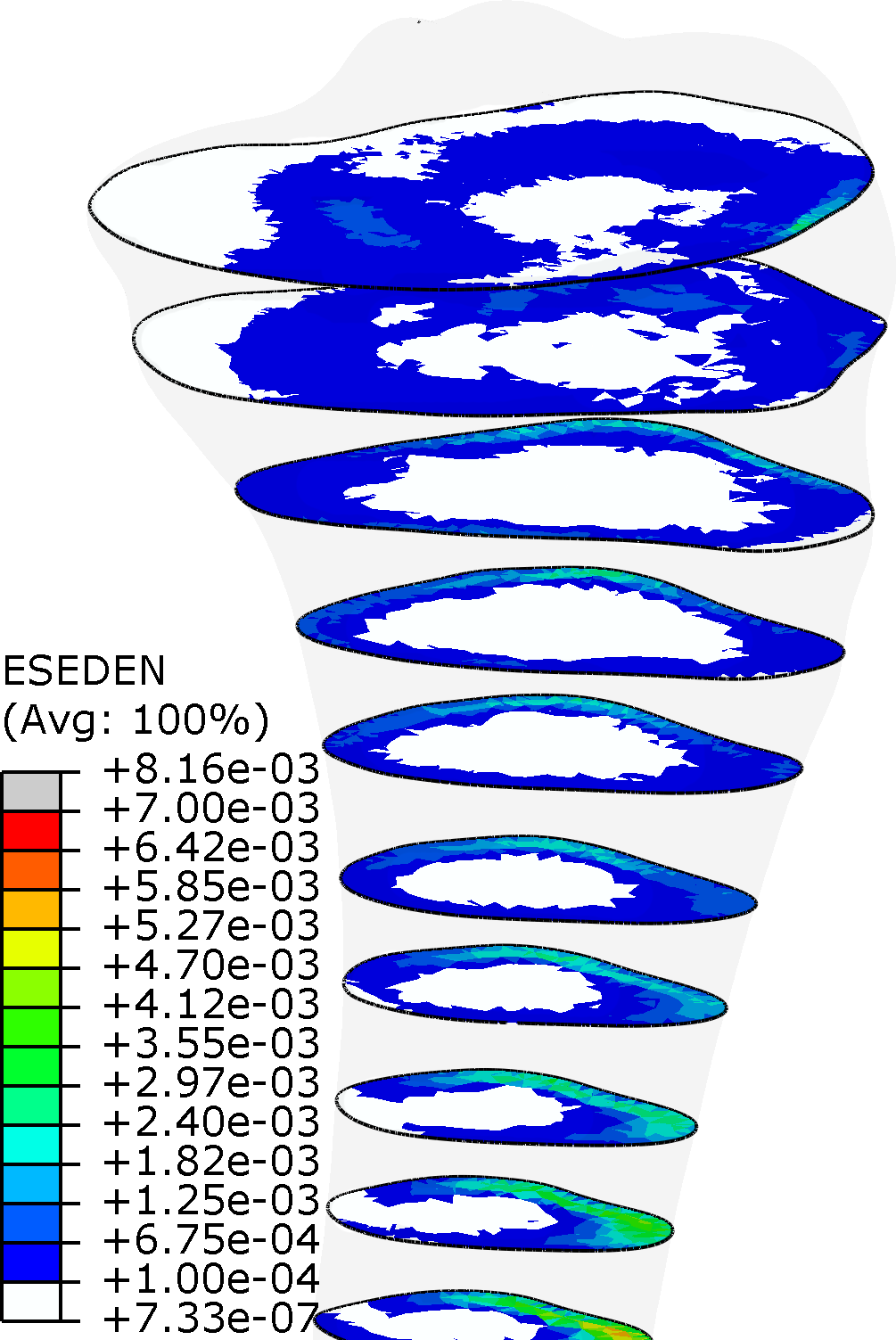}}
	    \\ 
		\hline 
	\end{tabular}
	\caption{{\bf Pre-surgery tibia:} heterogeneous distributions of Young's modulus (MPa) and of SED (MPa) with range adjustments to spongious and cortical bone. 
	\label{tab:YoungsMod-SEDpresurg-Tibia}}
\end{Table}

The images in Tab.~\ref{tab:YoungsMod-SEDpresurg-Tibia} display the heterogeneous Young's modulus distribution and the consecutive SED distribution in the tibia. Two different ranges are each chosen for visualization to account for the considerable differences between spongious and cortical bone in stiffness. As a consequence, SED distributions similarly exhibit a corresponding variation, which underpins that stiffness ''attracts'' bone loading in terms of SED.  

Notice that in the most proximal cross section (which approximately coincides with the surgical resection plane) tibial bone is mostly spongious attaining stiffness above 1500 MPa only at some spots. The considerably stiffer corticalis continuously forms a closed ring of increasing thickness in more distal cross sections. 

The distributions of Young's modulus and pre-surgery SED are the key reference for a proper interpretation of the post-surgery SED changes.

%------------------------------------------------------------------------------------------------------------------------   
\subsection{Postsurgery load transfer at the implant-bone interfaces}
\label{sec:load_transfer_at_resection_plane} 
 
\begin{Table}[htbp]
	\begin{minipage}{16.5cm}  
		\centering
		\renewcommand{\arraystretch}{1.2} 
		\begin{tabular}{lccccc}
			\hline
			& \multicolumn{4}{ c }{force transmission ratio [\%] \quad plate\,:\,stem  }                 \\
			\hline
			interface conditions     &       & implant &  \multicolumn{3}{ c }{extension length [mm]}     \\ 
%			\hline
			contact                  & cementation &  material &     75         &    40      &    5     \\ 
			\hline
			sticking friction        &  FullC   & CoCr  &  $17$:$83$   &  $27$:$73$  & $54$:$46$   \\
			                         &          & Ti    &  $23$:$77$   &  $30$:$70$  & $55$:$45$ \\
			                         &          & PE    &  $63$:$37$   &  $64$:$36$  & $74$:$26$ \\
			                          
			\cline{2-6}
			                         &  SurfC   & CoCr  &  $24$:$76$   &    $42$:$58$   & $61$:$39$   \\
			                         &          & Ti    &  $29$:$71$   &    $44$:$56$   & $62$:$38$   \\
			                         &          & PE    &  $81$:$19$   &    $81$:$19$   & $82$:$18$   \\
			\hline                         
			sliding friction cof=0.2 &  SurfC   & CoCr  &  $59$:$41$   &    $72$:$28$   & $56$:$44$   \\
			                         &          & Ti    &  $59$:$41$   &    $72$:$28$   & $56$:$44$   \\
			                         &          & PE    &  $71$:$29$   &    $81$:$19$   & $69$:$31$   \\	
			\cline{2-6}                          
	      \phantom{sliding friction} cof=0.0 &  SurfC   & CoCr &  $72$:$28$   &    $77$:$23$   & $62$:$38$    \\
	                                         &          & Ti   &  $72$:$28$   &    $77$:$23$   & $62$:$38$    \\          
			                         &          & PE   &  $79$:$21$   &    $84$:$16$   & $73$:$27$    \\
			\hline
%			fixation: cemented, (PMMA)        &       &  $55$:$45$   &    $30$:$70$     &    $23$:$77$     &    &      &        \\
%			\phantom{fixation:} silicon, $\nu=0.499$& &  $xx$:$yy$   &    $xx$:$yy$     &    $xx$:$yy$     &     &     &        \\
%			\hline
		\end{tabular}
		\newline 
	\end{minipage}
	\caption{Load partitioning for tibial plate-stem devices of various lengths, various  implant materials, and various interface conditions/fixations
		with FullC: full cementation, SurfC: cementation restricted to base plate. The applied force by the two femur condyles is symmetric, $|\bm F|=F_z= 2 \times 543$ N.. 	\label{tab:LoadPartitioning-PLate-vs-Stem2}}
\end{Table}   

Table~\ref{tab:LoadPartitioning-PLate-vs-Stem2} decomposes the force in $z$-direction of the coordinate system of Fig.~\ref{fig:Variants-of-endoprotheses} into its plate and stem parts. 
\begin{enumerate} 
	\item The larger the stiffness of the implant material along with the stiffness of the stem-bone interface, the smaller the plate:stem force ratio; for the stiffest case (CoCr for FullC) the ratio of 17:83 indicates a stem-dominated force transmission, for the most compliant case (PE for SurfC/sliding friction at cof=0.0) the ratio of 84:16 indicates a plate-dominated force transfer.
	\item Point (1) with its upper and lower bounds for force ratios spans a wide interval where a further analysis reveals the impacts of interface conditions and stem lengths.
	\begin{enumerate}
		\item For sliding friction/SurfC the plate transmits a considerably larger force portion than for sticking friction conditions. This is true for the metal-backed (MB) variants along with 75mm and 40mm stem extension lengths. For the 5\,mm case however, the difference between sticking and sliding friction mostly vanishes.
		\item For sticking friction conditions, in both the FullC and SurfC cases, it holds that the longer the stem, the larger the force portion transmitted by the stem. For sliding friction conditions, in both the cof=0.2 and cof=0.0 cases, the same is true, but only for stem extension lengths of 40\,mm and 75\,mm. In the 5mm case however in sliding friction, the force transmitted by the stem is throughout larger than for the longer stems.
		\item The plate:stem force ratio is invariant with respect to interface conditions (sticking or gliding) only for the short stem.
	\end{enumerate}
	\item The PE case is an exception to the above rules in that it is almost invariant with respect to stem length and interface conditions; the force portion transmitted through the plate is constant at around 80\%.
\end{enumerate}

Elementary mechanical reasoning suggests that the larger the force part mediated by the plate, the larger the tibial bone loading into proximal bone and more distal regions. Notice in this context that in natural, pre-surgery conditions the applied force is completely transmitted through that tibial plane which is chosen in surgery to be resection plane. {\color{black}In Sec.~\ref{sec:postsurgery_SED_differences} we show that the plate-to-stem force decomposition is in fact a reliable indicator for post-surgery to pre-surgery SED changes according to \eqref{eq:SED-difference-definition} in the proximal tibia.}  
\\[2mm] 
{\color{black} In reality, the cement usually balloons out at the tip region of the stem, frequently contacting the posterior cortex and undoubtedly transmitting loads in the area. In that case the axial force portion transmitted by the stem is expected to rise. In the present analysis however, where a 40 mm stem along with a balloon of 10 mm diameter is considered, force transmission remains unaltered Tab.~\ref{tab:LoadPartitioning-StemTipConditions}, which indicates that a stiff bridge to the cortex was not established. For the uncemented stem the tip conditions --hollow versus spongy bone-- similarly have minor influence on force transmission}. 

\begin{Table}[htbp]
	\begin{minipage}{16.5cm}  
	\centering
	\renewcommand{\arraystretch}{0.8}
	{\begin{tabular}{lccccc}
			\toprule
			& & & \multicolumn{3}{ c }{force transmission ratio [\%] \quad plate\,:\,stem  }                 \\
			\midrule
			interface conditions     &       & implant &  \multicolumn{3}{ c }{\color{black}stem tip conditions}     \\ 
			\cmidrule{4-6}
			contact                  & cementation & material & hollow & spongy bone & cement balloon \\ 
			\midrule
			sliding friction cof=0.2 &  SurfC  & Ti  & $75$:$25$ & $72$:$28$ & -- \\
			\midrule
			sticking friction        &  FullC  & Ti  & -- & $30$:$70$ & $30$:$70$ \\
			\bottomrule
	\end{tabular}}
	\newline 
	\end{minipage}
	\caption{\color{black}Load partitioning for tibial plate-stem devices for Ti-alloy, stem length 40 mm, surface cementation with dependency of stem tip conditions. Applied forces as in Tab. \ref{tab:LoadPartitioning-PLate-vs-Stem2}.\label{tab:LoadPartitioning-StemTipConditions}}
	
\end{Table}

\subsection{Postsurgery SED differences}
\label{sec:postsurgery_SED_differences} 
 
\begin{Table}[htbp]
	\centering
	% \renewcommand{\arraystretch}{1.2}
%	\begin{tabular}{| c | c c c c |}
		%
 \begin{tabular}{| >{\centering\arraybackslash} m{10mm} | >{\centering\arraybackslash} m{3.2cm} >{\centering\arraybackslash} m{3.2cm} >{\centering\arraybackslash} m{3.2cm} >{\centering\arraybackslash} m{3.4cm} |}
		\hline  
		& & & & \\[-2mm]
		stem   & Full cementation &  \multicolumn{3}{c|}{ ------------------ Surface cementation ------------------ }  \\[2mm] 
		length & \multicolumn{2}{c}{ ------ sticking friction conditions ------}  & 
		\multicolumn{2}{c|}{ ------ {\bf sliding friction conditions} ------}   \\[1mm] 
		(mm)   & thickness 2mm & osseointegration  & cof=0.2 & cof=0.0 \\ 
		\hline
		& & & & \\[-3mm]
		75  & \includegraphics[height=6.0cm, angle=0]{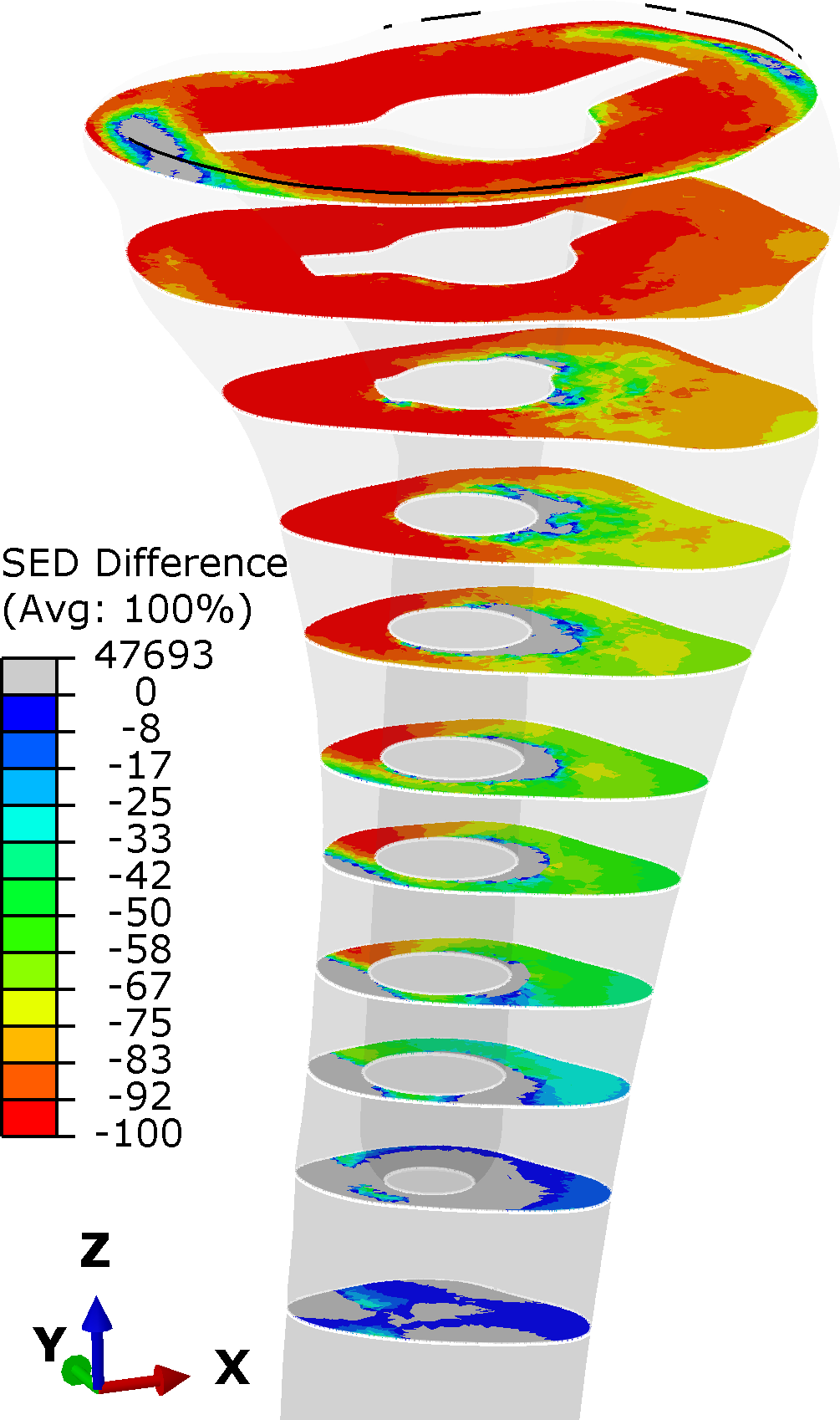}
		& \includegraphics[height=6.0cm, angle=0]{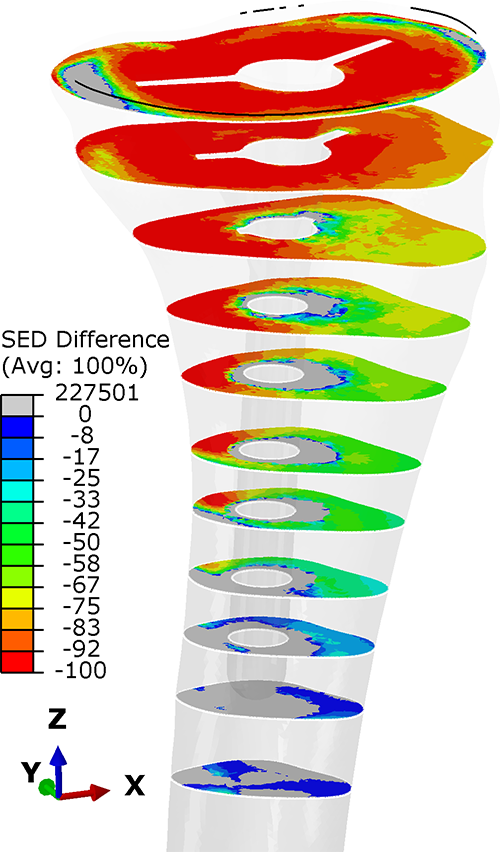} 
		& \includegraphics[height=6.0cm, angle=0]{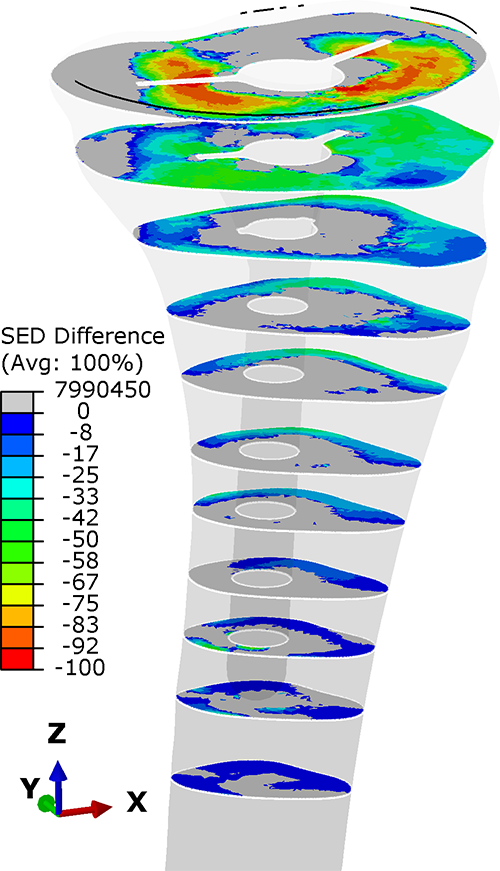}
		& \includegraphics[height=6.0cm, angle=0]{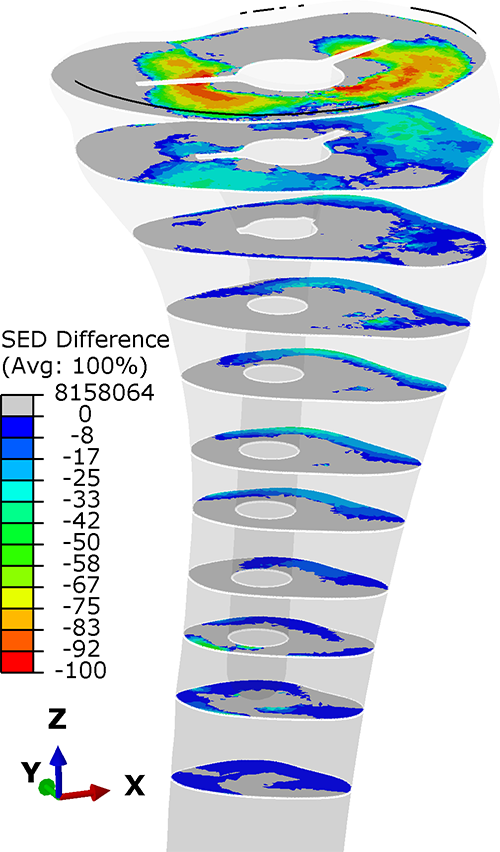} \\[1mm]
		& 23:77 & 29:71 & 59:41 & 72:28 \\
		\hline  
		& & & & \\[-3mm]
		40  & \includegraphics[height=4.8cm, angle=0]{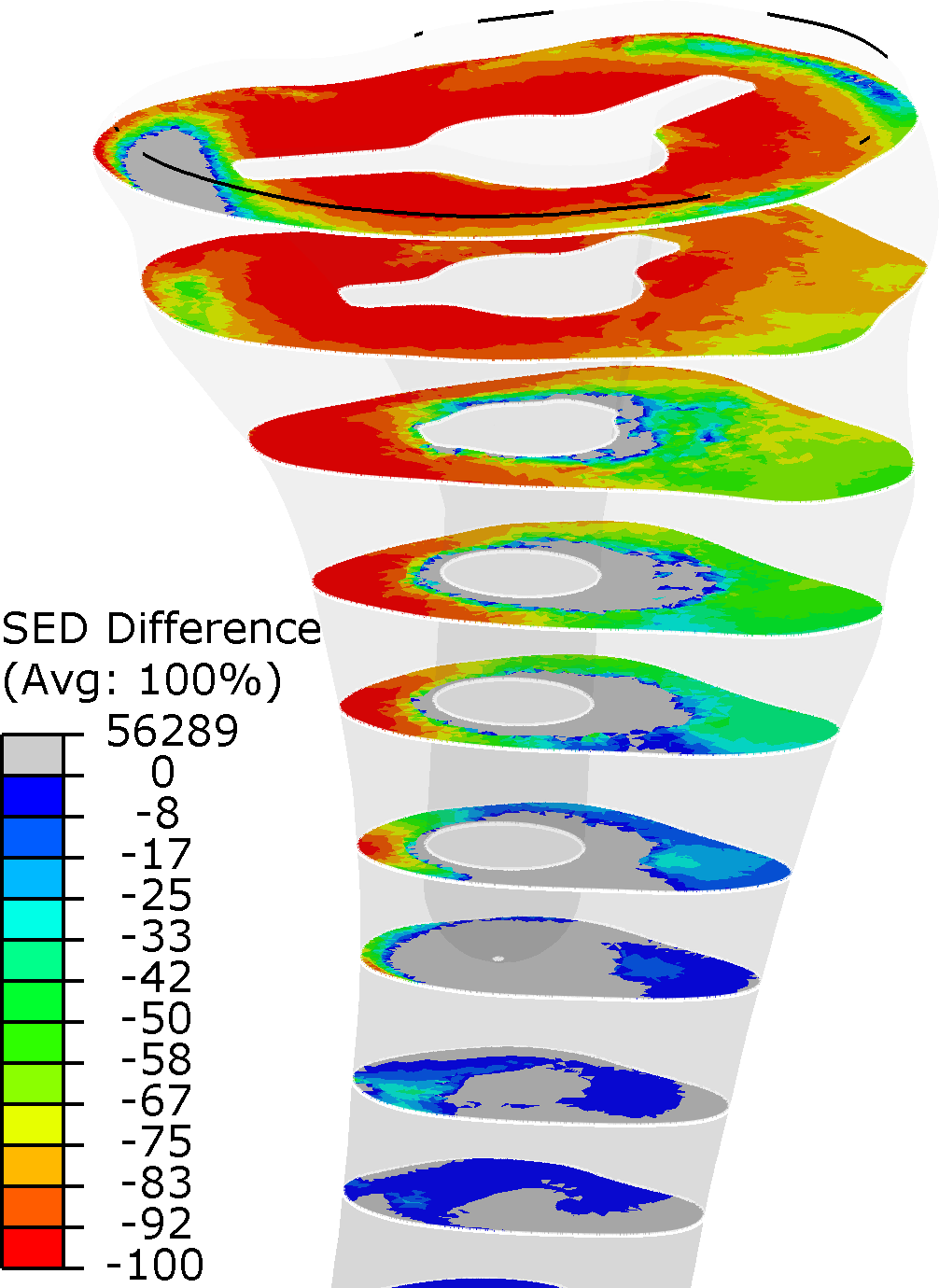}
		& \includegraphics[height=4.8cm, angle=0]{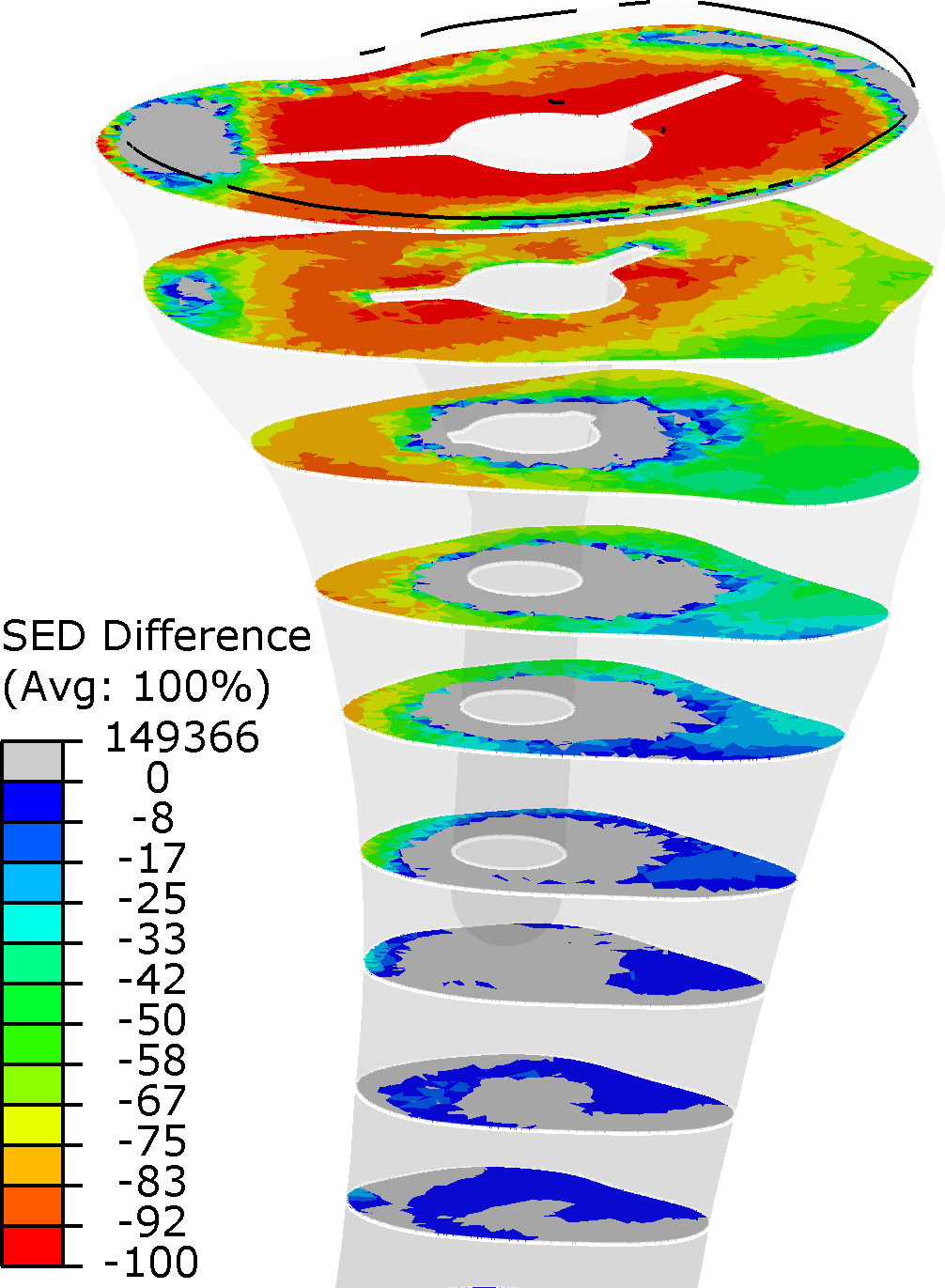}
		& \includegraphics[height=4.8cm, angle=0]{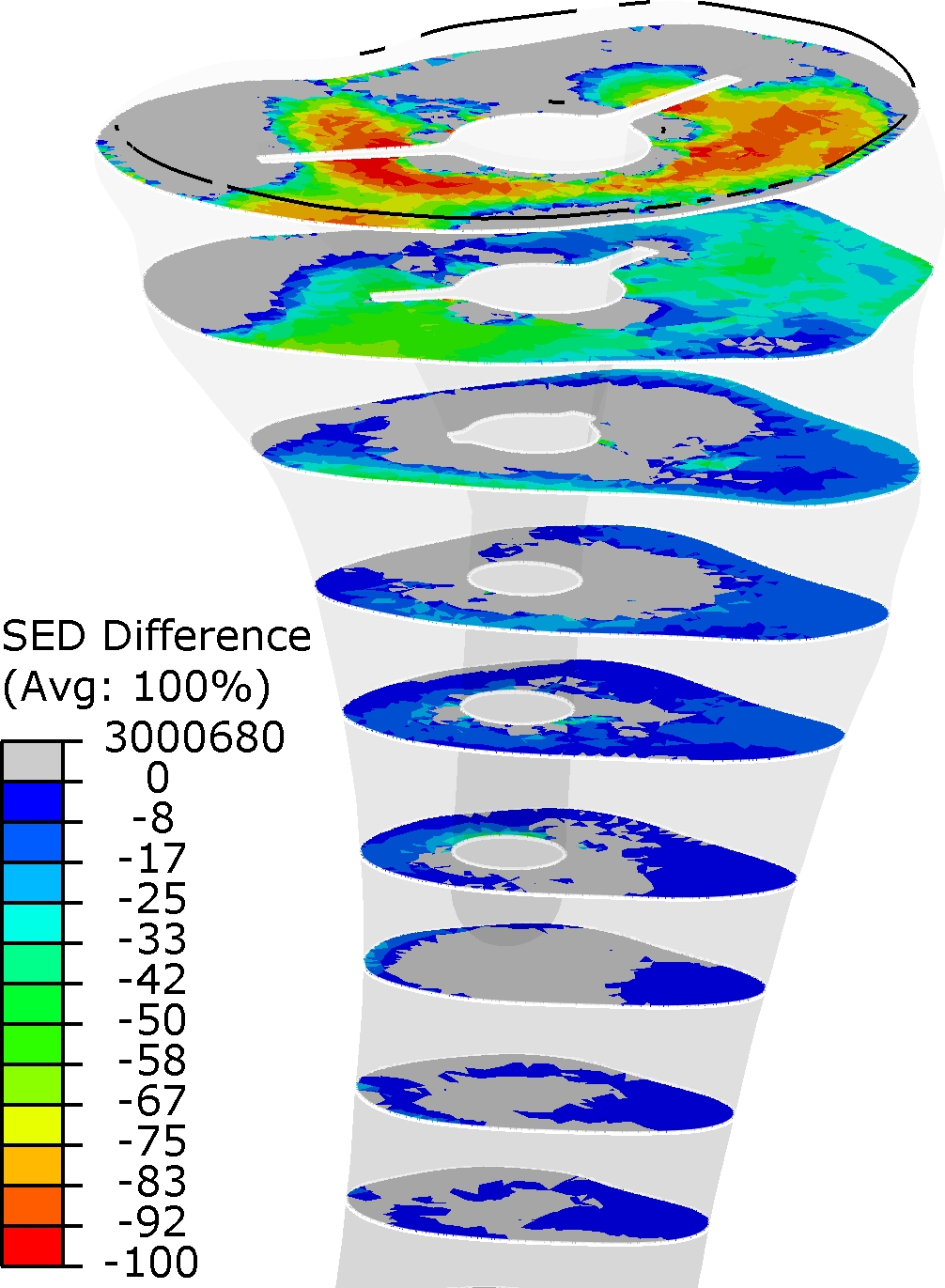}
		& \includegraphics[height=4.8cm, angle=0]{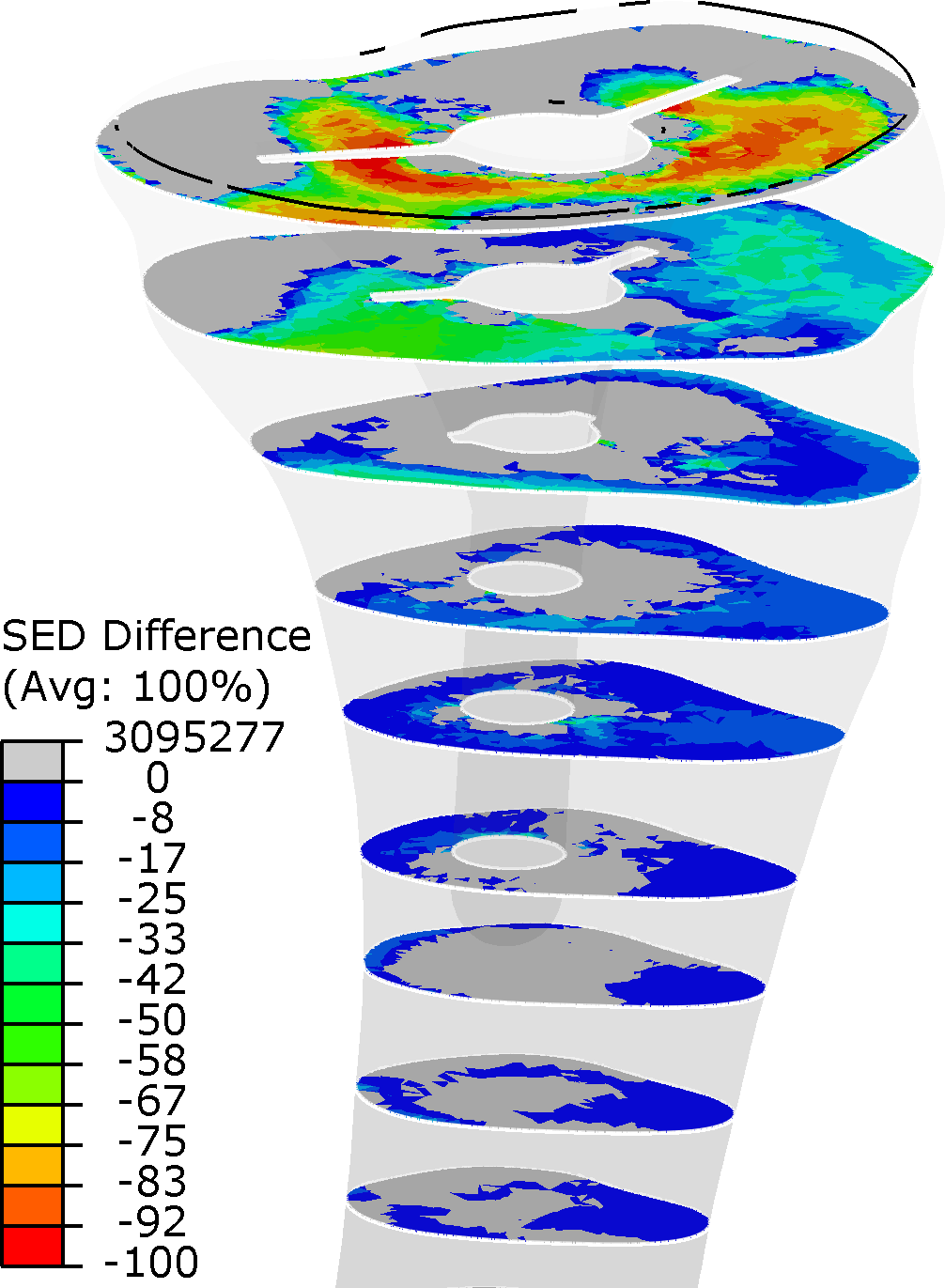} \\[1mm]
		& 30:70 & 44:56 & 72:28 & 77:23  \\
		\hline  
		& & & & \\[-3mm]
		5   & \includegraphics[height=4.8cm, angle=0]{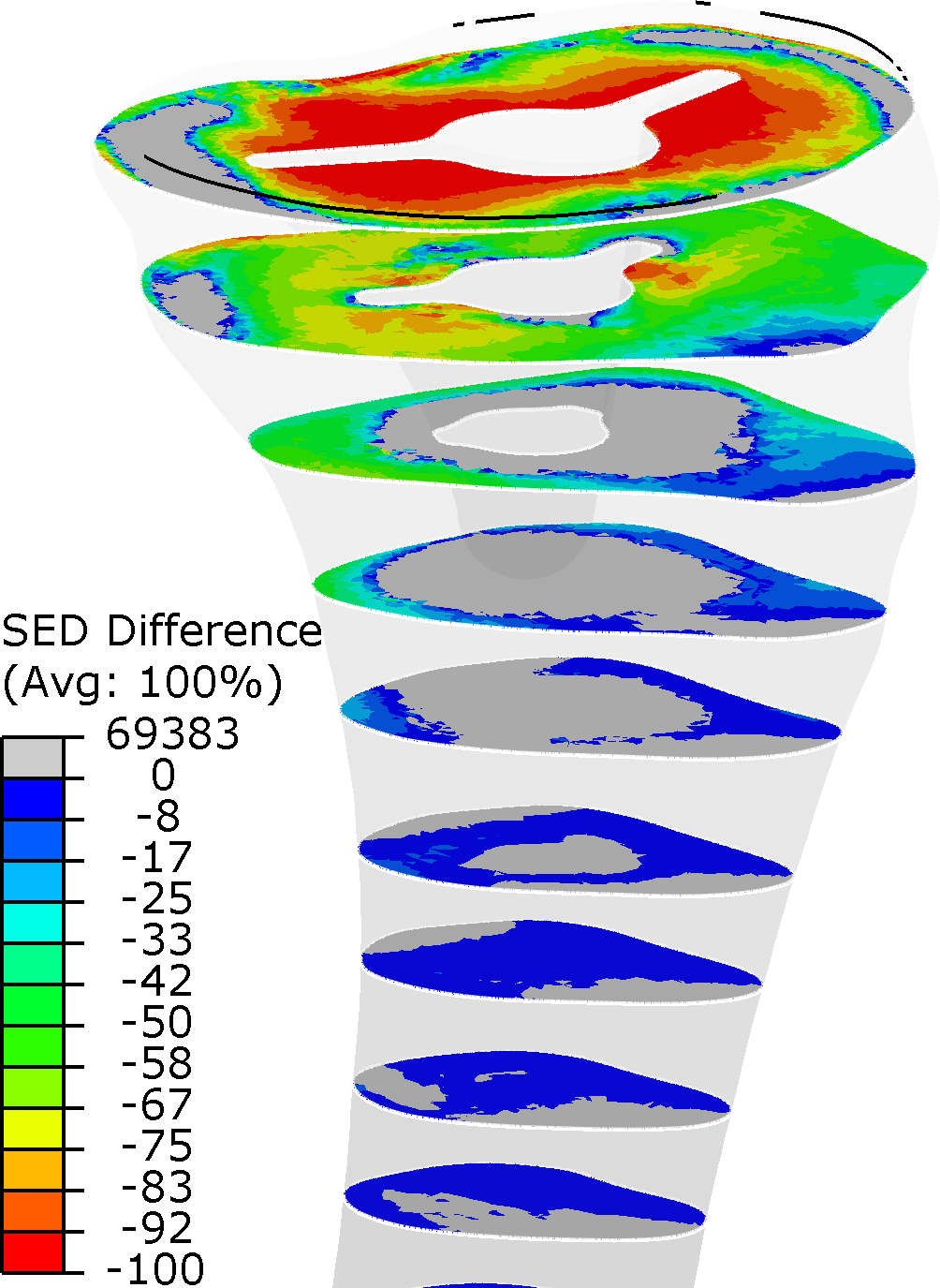}
		& \includegraphics[height=4.8cm, angle=0]{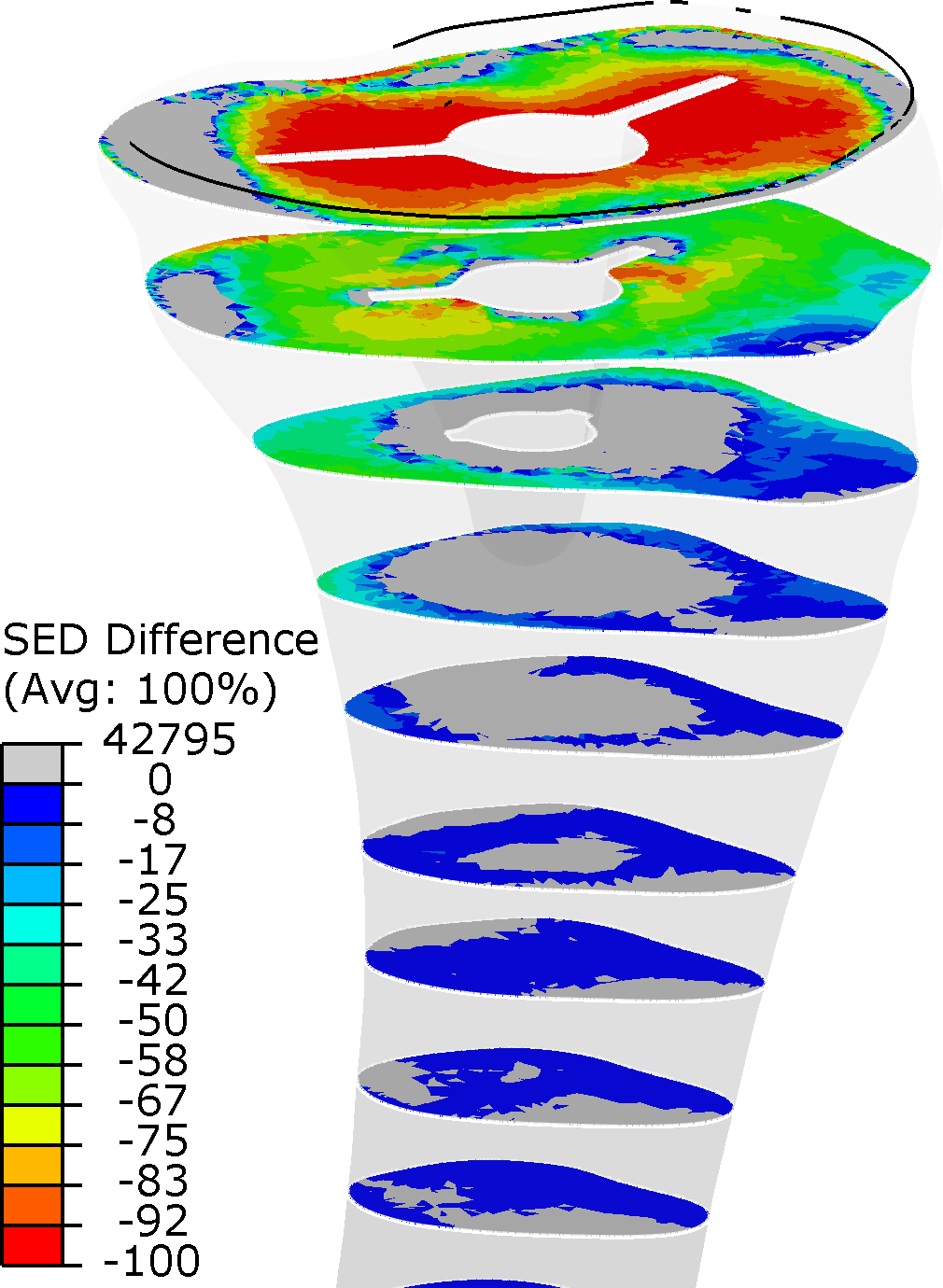}
		& \includegraphics[height=4.8cm, angle=0]{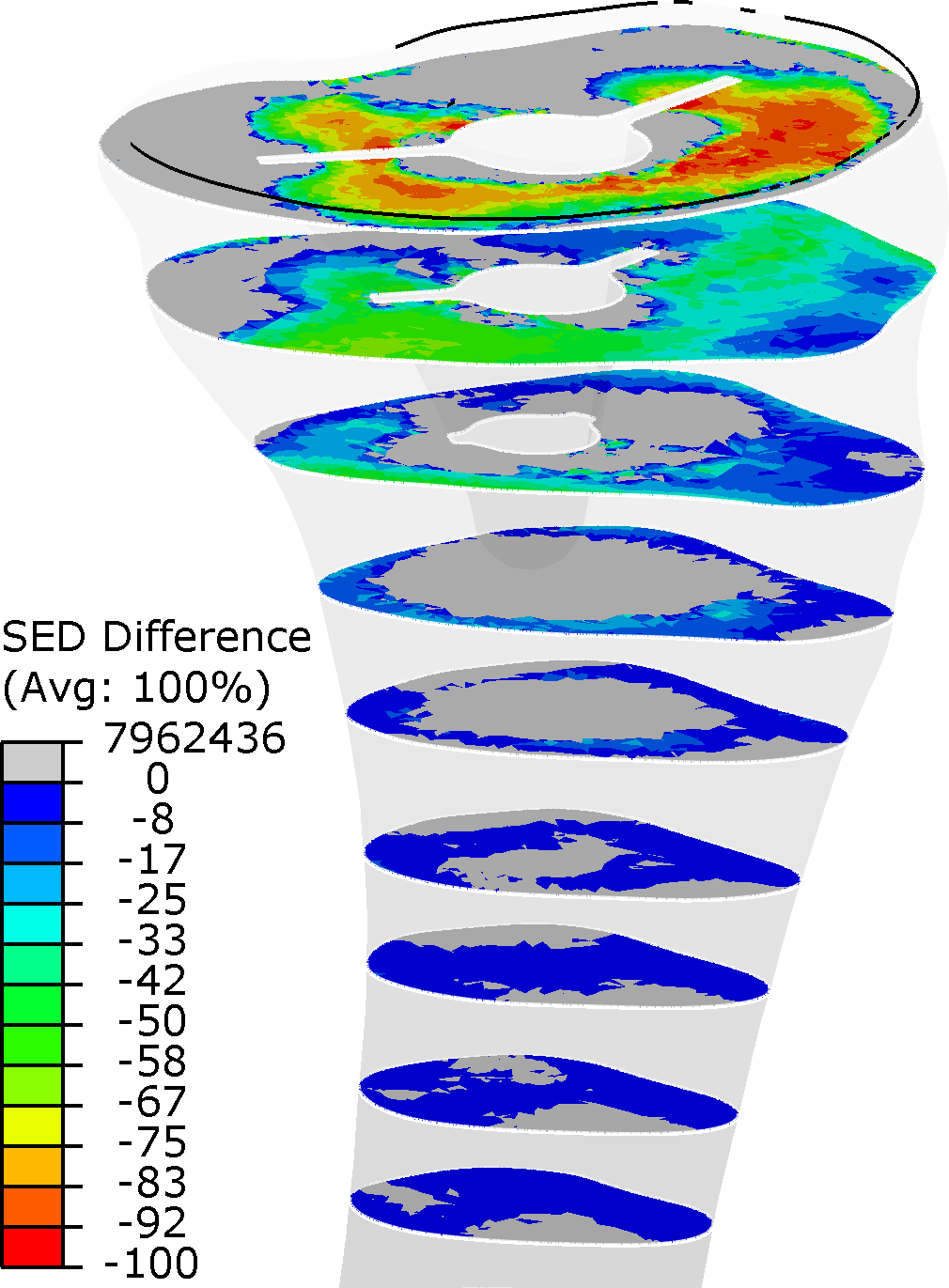}
		& \includegraphics[height=4.8cm, angle=0]{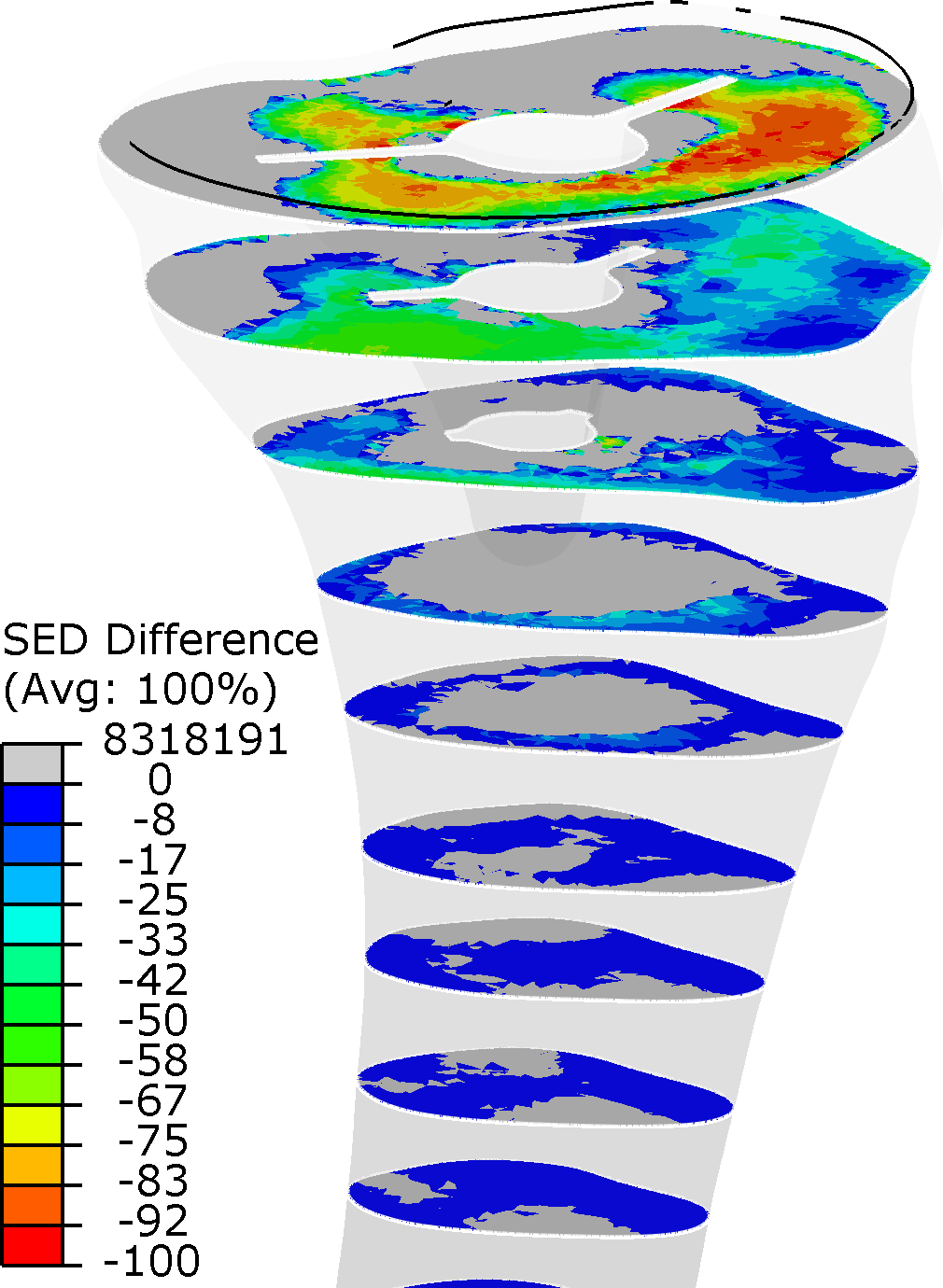} \\[2mm]
		& 55:45 & 62:38 & 56:44 & 62:38 \\
		\hline 
	\end{tabular}
	\caption{{{\bf Post- to pre-surgery SED differences [\%] and plate:stem force decomposition [\%]}} for Ti-alloy implant of various stem lengths and different interface conditions.}
	\label{fig:SEDdiff-Ti-variousBCs-variousStemLength-table}
\end{Table}
  
The impact of interface conditions and stem extension length on SED differences is visualized in Table~\ref{fig:SEDdiff-Ti-variousBCs-variousStemLength-table} for the Ti-alloy implant. 
\begin{itemize}
	\item The lower the shear stiffness of the stem-bone interface, (decreasing from left to right in 
	Table.~\ref{fig:SEDdiff-Ti-variousBCs-variousStemLength-table}) the smaller the post-surgery SED loss both in intensity and spatial extension.
	\item The longer the stem extension in the case of sticking friction, the stronger and more distally extended is the SED loss, a consequence which is even more pronounced for the fully cemented fixation than for the case of press-fit/osseointegration at the stem.
	For sliding friction conditions however, the influence of stem length on post-surgery SED loss is confined to the proximal tibia, but distally very limited.
	\item The newly introduced force decomposition into plate and stem parts is a definite, reliable measure for the intensity and spatial extension of post-surgery SED reduction; if primarily the plate transmits the axial force, SED reduction is confined in magnitude and to small regions in most proximal cross sections, for the case of stem-mediated force transmission the opposite is true.
\end{itemize}
%---------------------------------------------     

\begin{Table}[htbp]
\begin{tabular}{| >{\centering\arraybackslash} m{18mm} | >{\centering\arraybackslash} m{4.0cm} >{\centering\arraybackslash} m{4.0cm} >{\centering\arraybackslash} m{4.0cm}  |}  
\hline 
& & & \\[-2mm]
% length [mm] & cemented fixation thickness: 2mm &  sticking friction & sliding friction cof=0.2 &  sliding friction cof=0.0 \\
interface   & CoCr alloy       &  Ti-alloy    &  all-PE       \\[0mm]
conditions  & $E$=220\,000 MPa & $E$=110\,000 MPa &  $E$=1200 MPa \\[2mm]
\hline
& & & \\[-2mm]
sticking friction & \includegraphics[height=6.5cm, angle=0]{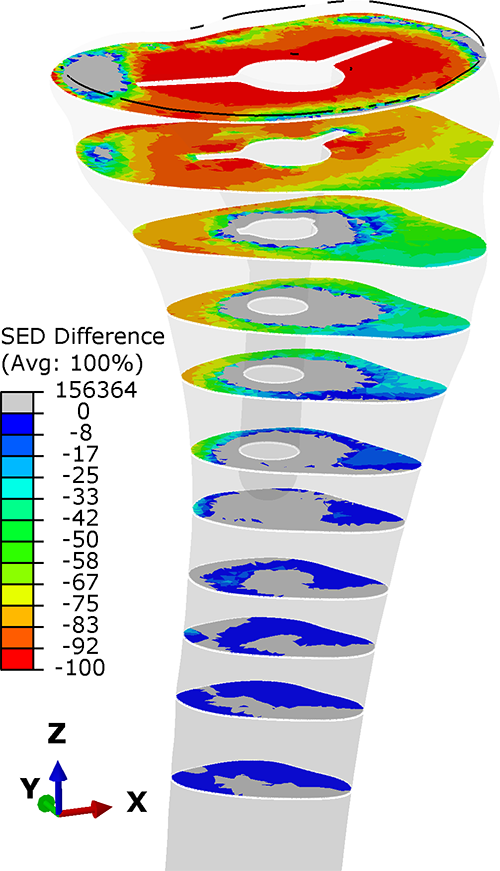}
          & \includegraphics[height=6.5cm, angle=0]{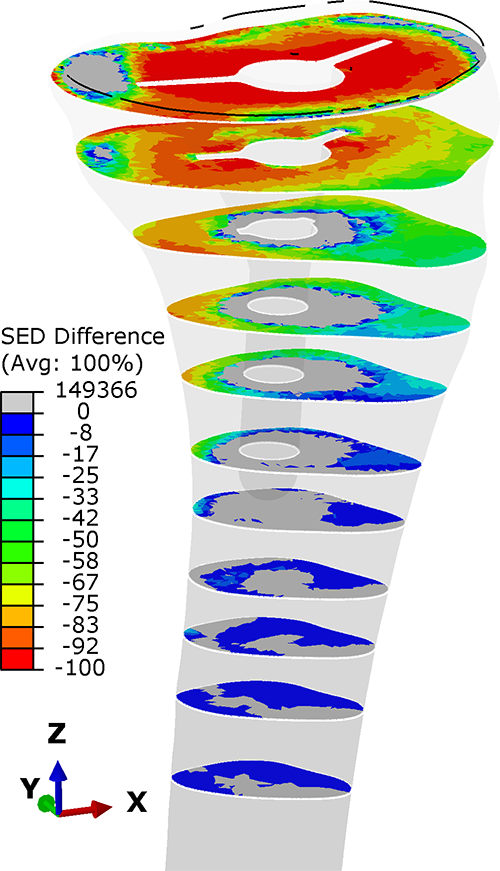} 
          & \includegraphics[height=6.5cm, angle=0]{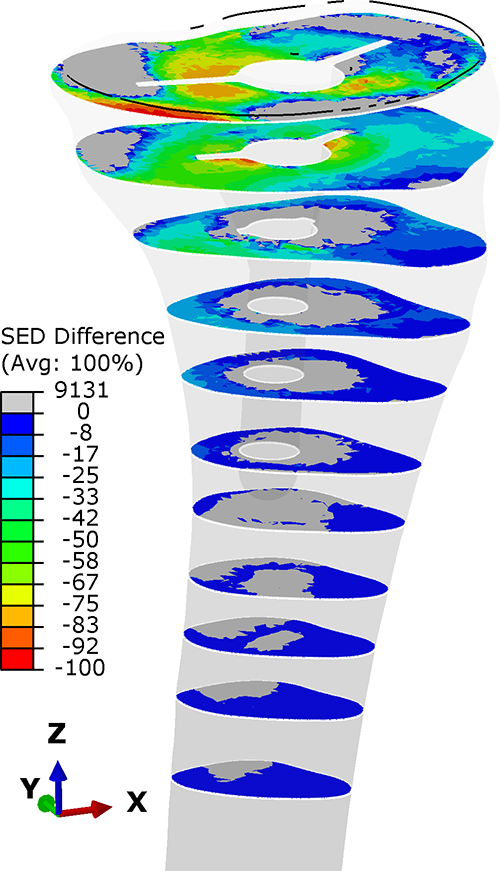} \\[1mm]
          &  42:58 & 44:56  & 81:19 \\
\hline  
& & & \\[-2mm]
sliding friction cof=0.2  & \includegraphics[height=6.5cm, angle=0]{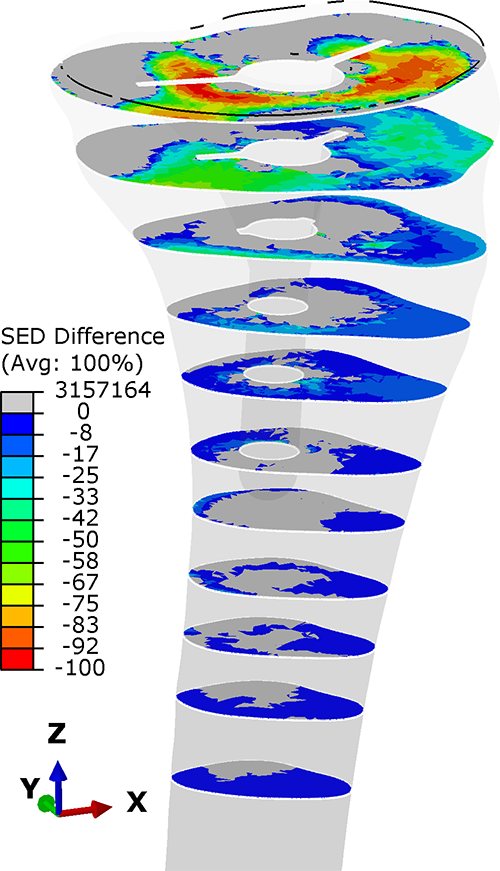}
          & \includegraphics[height=6.5cm, angle=0]{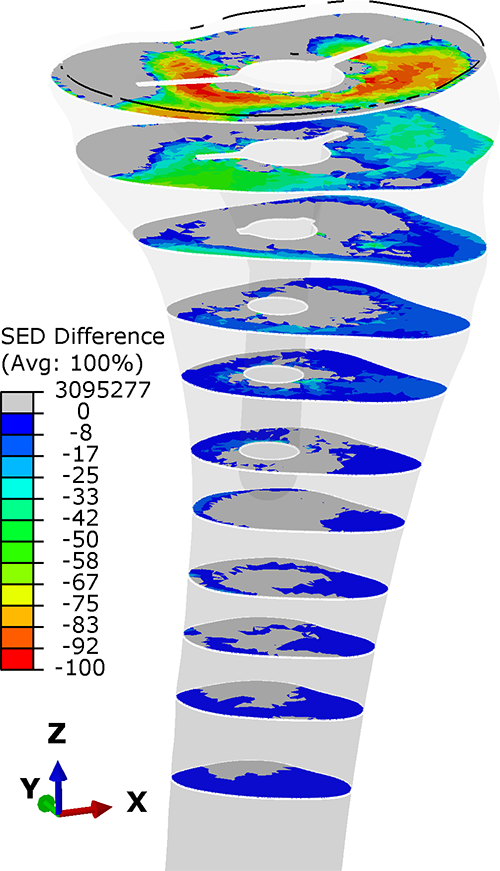}
          & \includegraphics[height=6.5cm, angle=0]{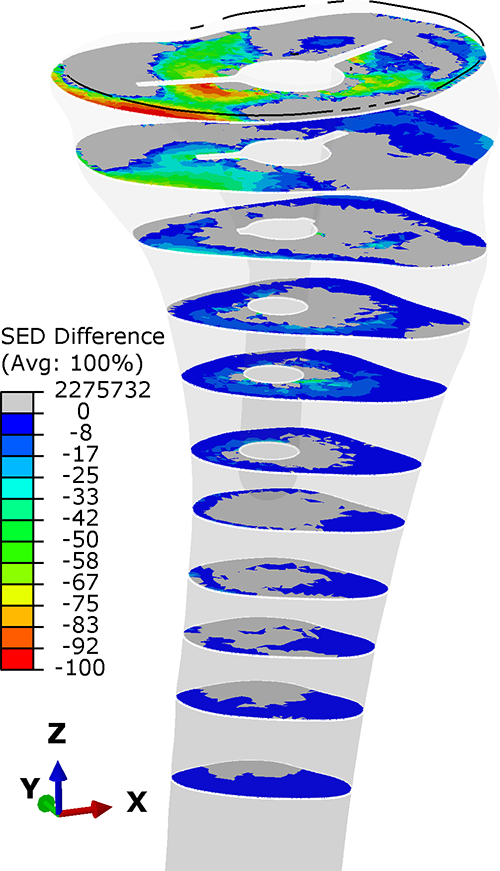} \\[1mm]
          & 77:23 & 77:23 & 84:16 \\
\hline 
   
\end{tabular}
\caption{\color{black}{{\bf Post- to presurgery SED differences [\%] and plate:stem force decomposition [\%]}} for various implant materials and interface conditions at stem length of 40mm and cementless stem fixation.}
\label{fig:SEDdiff-40-VariousMaterials-VariousInterfaceCond-tab}
\end{Table}
 
The impact of the implant material on SED-changes is shown in Table~\ref{fig:SEDdiff-40-VariousMaterials-VariousInterfaceCond-tab}. 
The comparison of the CoCr-case with the Ti-alloy case indicates that stiffness change of the implant material, here roughly 2:1, has very little influence on SED changes compared to pre-surgery conditions, which is true for both sticking and sliding friction conditions at low friction coefficients. It is only the sticking friction case, where the drastic Young's modulus reduction from CoCr/Ti (210/110 GPa) to PE (1.2 GPa) material (Young's modulus ratio Ti-alloy to PE: 91) does considerably avoid SED loss. For sliding friction however, the improvement of the fully PE implant compared to the much stiffer materials CoCr and Ti is minor. 

If the stiffnesses of bone and implant are decoupled by sliding friction conditions, the stiffness of the implant does not matter; the all-PE variant exhibits almost the same SED-differences as the much stiffer Ti-alloy does. {\color{black}Only in the case of sticking friction realized by full cementation or osseointegration, material stiffness has a stronger impact on stress shielding.}

\section{Discussion}

\subsection{What is and what drives stress shielding? How can it be overcome?}

Stress shielding refers to the reduction in bone density caused by a reduction of physiological  stress from the bone by a stiff, metallic implant. This is because of bone's adaptivity to remodel in response to the loads referred to as Wolff's law. As a consequence, a decrease of loading on bone decreases bone density and stiffness because of the lowered stimulus for maintaining the existing bone density.

While none of these statements is wrong, the present work suggests a distinguished analysis of stress shielding with respect to its roots.  

The point of departure is the finding that the force transmitted by the tibial plate is relatively small, if the stem is rigidly connected to bone, either by press-fit/osseointegration in cementless fixation (referred to as surface cementation) or by cemented fixation (referred to as full cementation). In that case, it is the stem which transmits by shear stresses the majority of the applied load in axial direction into bone. As a consequence, the proximal regions in the tibia suffer from considerable SED reduction; in view of the necessity for vital bone cells to get fed by loading it can be seen as a SED-starvation. This effect expands to more distal tibia parts the longer the tibial stem extension. 

The present work shows that the key aspect to overcome stress shielding in tibial bone is the mechanical activation of the tibia plate for the implant-to-bone force transfer. The activation is realized by resolving the sticking connection between implant stem and bone\footnote{The plate remains fixed to bone by cement (PMMA).}, which implies that the cemented interface shall be avoided. For the case of sliding friction in the regime of low coefficients of friction (cof=0.2) SED reduction is bounded to modest magnitude almost everywhere, if present at all. 

The above described effects follow the regime of decreasing stiffnesses: FullC $>$ SurfC/osseointegration $>$ SurfC/sliding friction, cof=0.2 $>$ sliding friction, cof=0.0. The comparison of FullC with SurfC for sticking friction (hence the two stiffest cases in the above hierarchy) was analyzed in \cite{Cawley.2013}, where the authors find a stress reduction under the tibial baseplate for both cases, but stronger for the full cementation.\\[2mm]  
%--------------------------------------------------------------------
{\bf Impact of implant-to-bone stiffness mismatch.} Even if the considerable stiffness mismatch between implant material and bone persists (140:1 for the CrMo case, 73:1 for the Ti-alloy) stress shielding is overcome for compliant stem-to-bone interface conditions. 
Remarkably, in terms of SED-preservation the fully polyethylene implant ($E$=1200 MPa) rigidly connected to bone does not excel the Ti-based implant, if the latter exhibits a smooth stem surface avoiding force transmission by shear. 

The notion of stress-shielding as an inevitable consequence of the implant-bone stiffness mismatch is misleading, since stress shielding becomes effective only for a rigid  stem-to-bone connection, by cementation or osseointegration. Among great many descriptions of the mechanical sources of stress-shielding, Gefen's definition is one of the very few precise ones \cite{Gefen.2002}.

The smaller the stiffness of the stem-to-bone connection, the lower the SED-reduction. This is a simple consequence of a reduced force transfer through the plate for stiffer shear support along the stem. The simulation results underpin this hierarchy. The use of short-keeled cemented tibial components however is no generally best solution, since an increased risk for aseptic loosening was reported \cite{Ries.2013}.

\subsection{Post-TKA excess SED, source and consequences}

In all cases considered in Tables \ref{fig:SEDdiff-Ti-variousBCs-variousStemLength-table} and \ref{fig:SEDdiff-40-VariousMaterials-VariousInterfaceCond-tab} some regions indicate a drastic post-surgery SED increase, which calls for a mechanical explanation addressing whether it is of harm.  To put things into perspective, the images in Figure~\ref{fig:Postsurgery-Excess-SED} (a) and (b) underpin that despite the huge maximal values of SED increase, excess SED is bounded both in magnitude and spatial extension. Moreover, it becomes transparent that along the stem the SED increase is in the soft spongious core of the tibia, where pre-surgery tibia exhibits  low SED values (c) due to low stiffness (d) such that a drastic post-surgery increase does not necessarily exceed the bone strength.
\begin{Figure}[htbp]
	\begin{minipage}{16.5cm}  
		\centering
		\subfigure[]{\includegraphics[height=6.4cm, angle=0]{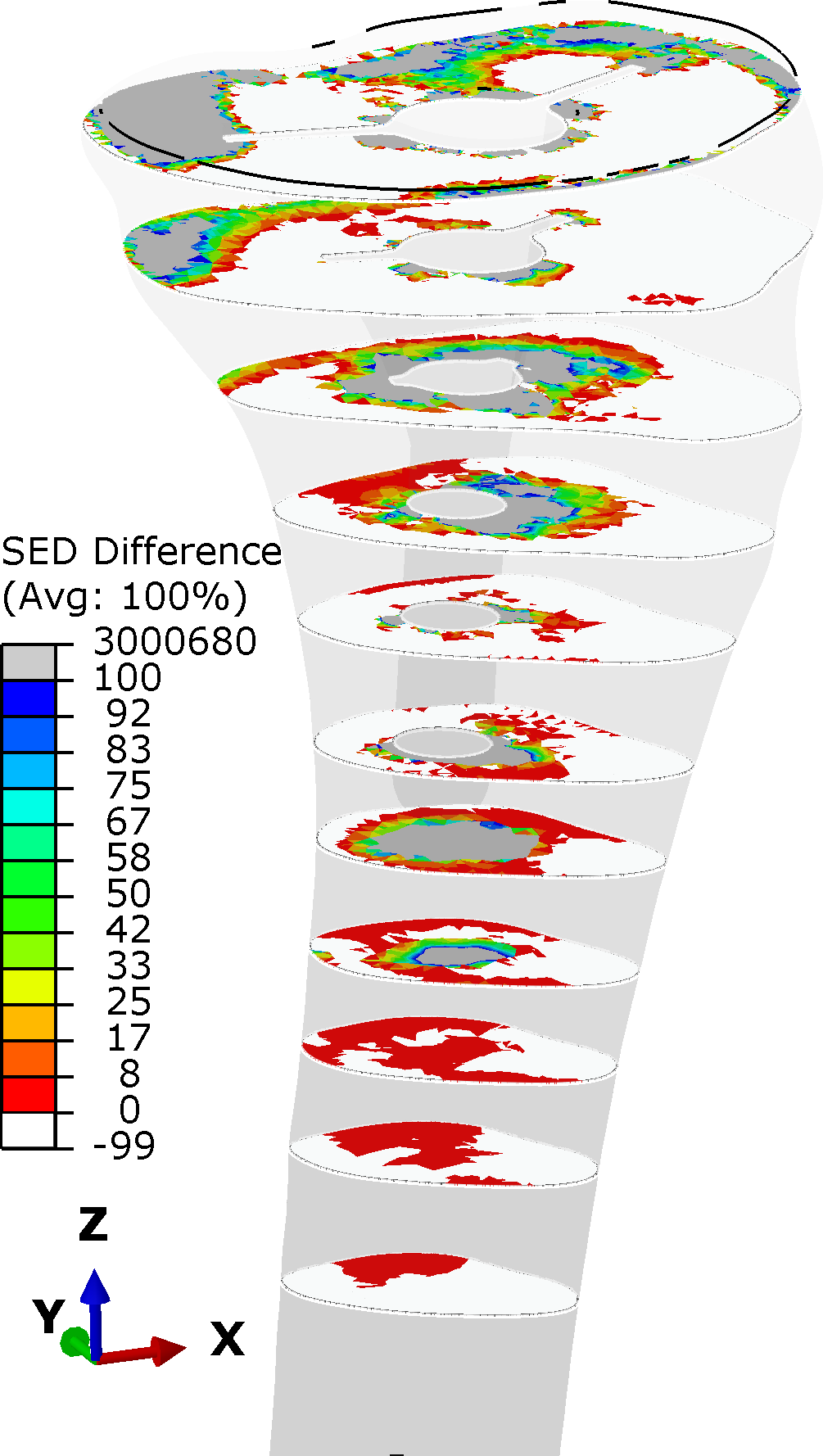}}  \hspace*{-2mm}
		\subfigure[]{\includegraphics[height=6.4cm, angle=0]{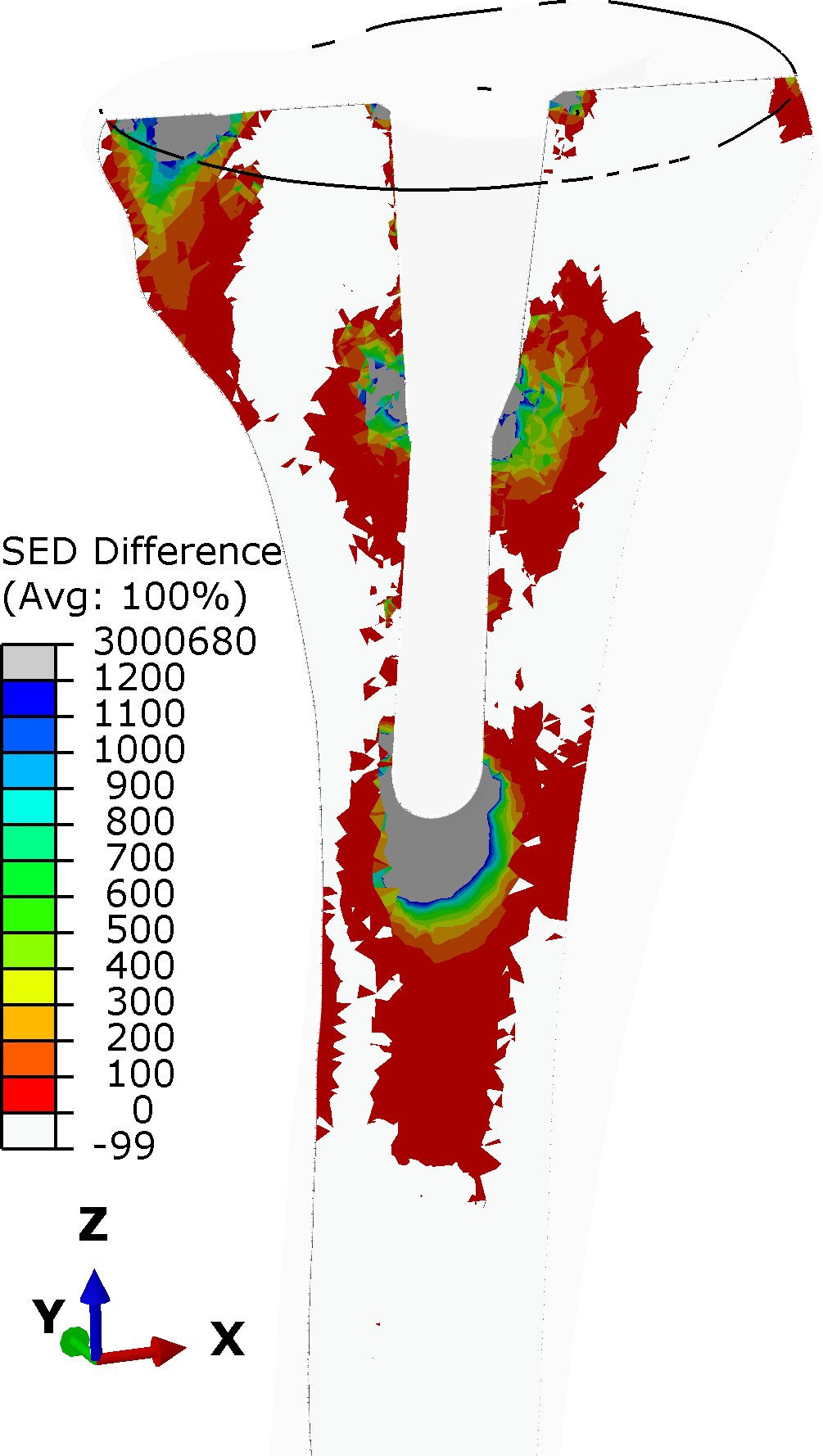}} \hspace*{-2mm} 
		\subfigure[]{\includegraphics[height=6.4cm, angle=0]{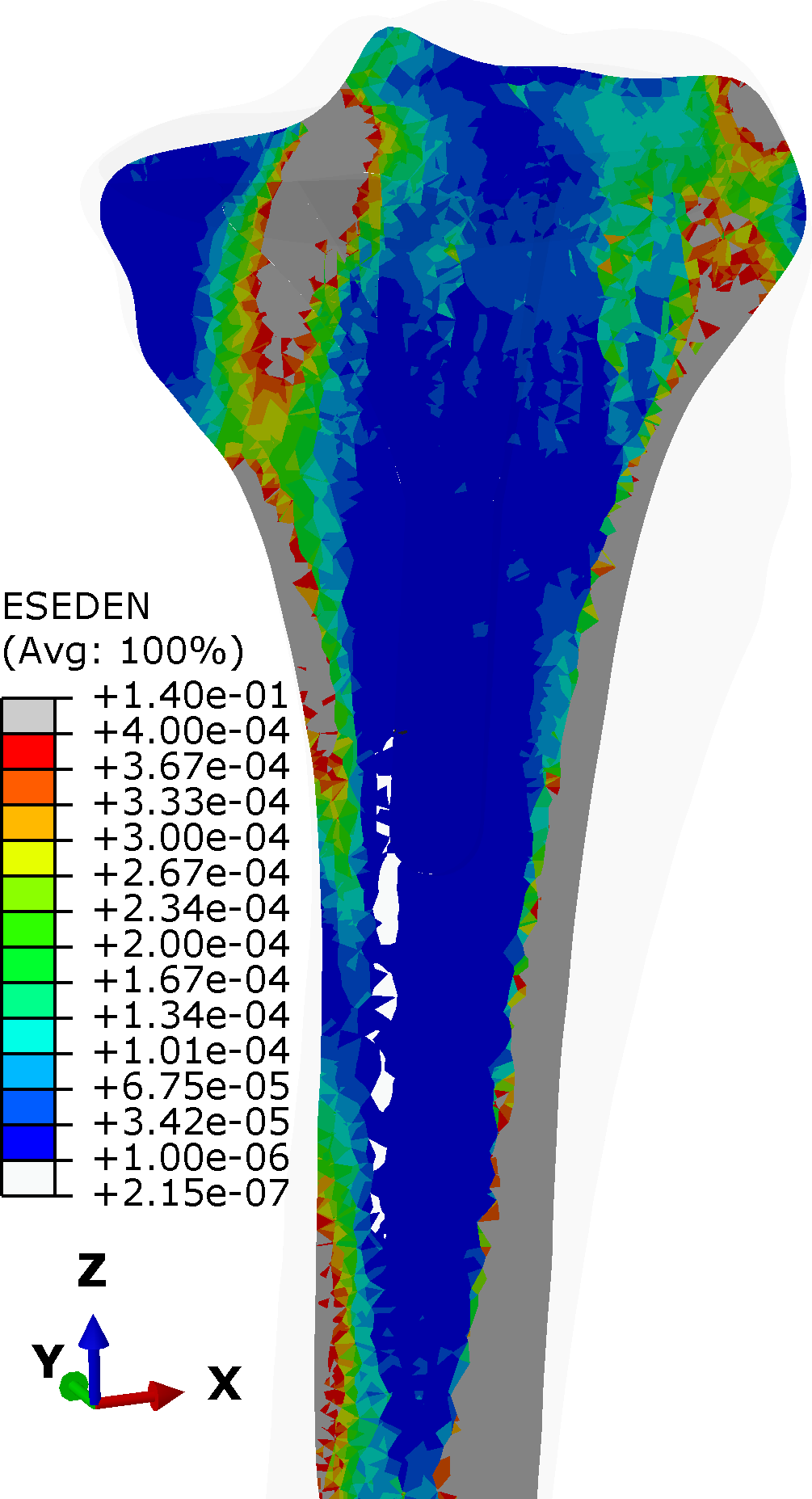}} \hspace*{-2mm} 
		\subfigure[]{\includegraphics[height=6.4cm, angle=0]{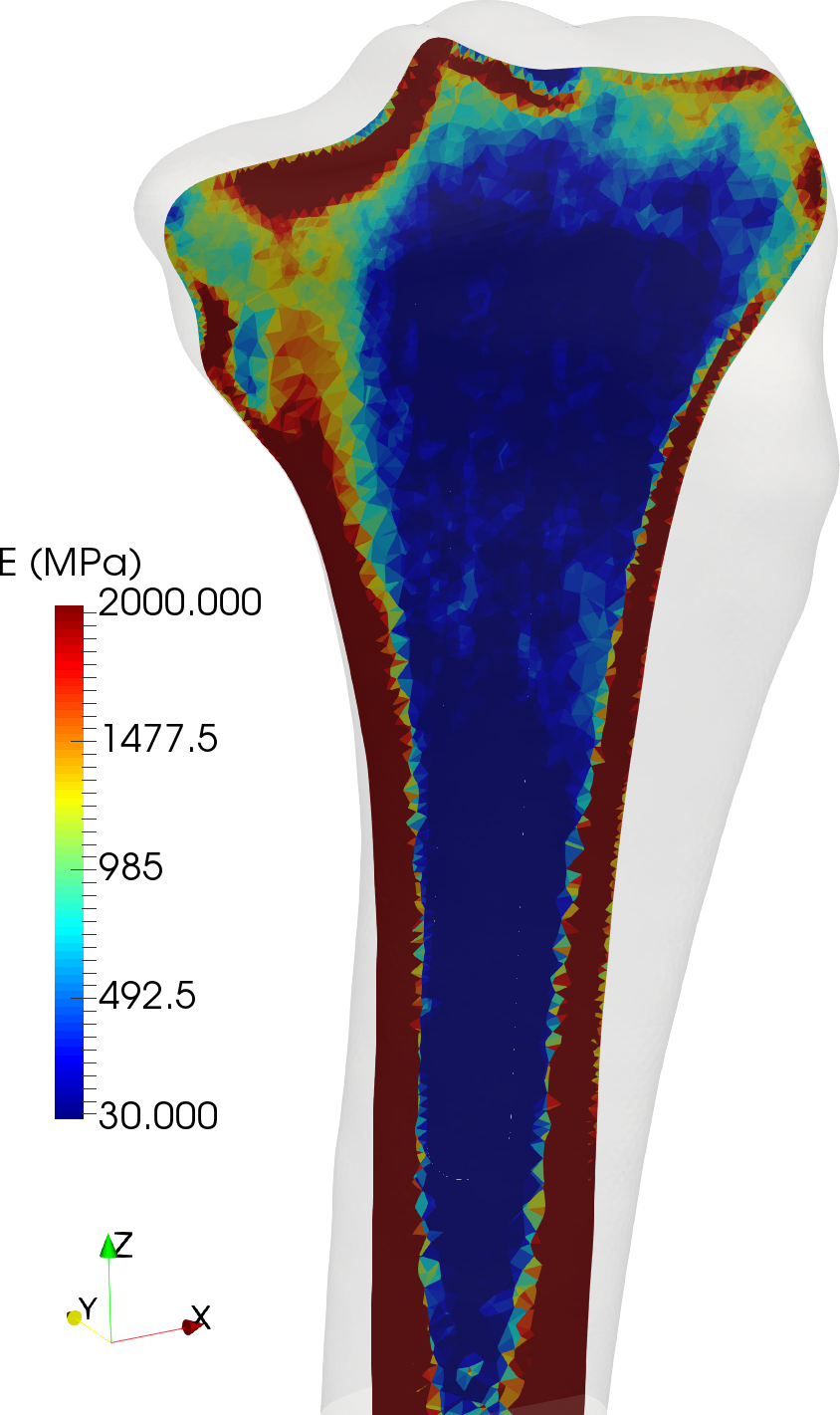}} \hspace*{-2mm}
		%\subfigure[]{\includegraphics[height=4.5cm, angle=0]{surface-coverage_MD_cut.eps}} 
	\end{minipage}
	\caption{{\bf Post-surgery excess SED:} SED increase in (a) planes perpendicular to the $z$-axis and (b) in a cross section along the stem axis, (c) pre-surgery SED distribution adapted to spongious tibia and (d) Young's modulus distribution.
		\label{fig:Postsurgery-Excess-SED}}
\end{Figure}
 
\subsection{Realization} 
\label{subsec:realization}
The novel fixation concept can be realized without considerable changes in implant geometry and surgery techniques. The surface of the stem must be sufficiently smooth. For realizing a low-friction regime at the implant-bone interface, a protective thin film of a biocompatible Diamond-Like Carbon (DLC) is favorable and well-established for its excellent bio- and hemocompatibility \cite{Robertson.2002}. Friction coefficients generally in the range of 0.05-0.2 have been reported for DLC \cite{Grill.1997} which are considerably lower than oxidized Ti-alloy surfaces \cite{RamosSaenz.2010}.

{\color{black} In classification point of view, the proposed concept belongs to the case of surface cementation along with a press-fitting of the tibial stem in the tibial physis at vanishing press fit and very small friction coefficient. This implies decreased bone loss on revision compared to a stem rigidly connected to bone either by cementation or osseointegration.} 

%Argue that sufficient implant stability is achieved with decreased metaphyseal bone loss on revision and without the potential for stress shielding thought to be associated with the cemented stem, see \citep{Peters.2003}. They show by biomechanical cadaver study, that no difference in micromotion of either tibial component implanted with surface or full cementation. 

\subsection{Stability in view of micromotion} {\color{black} Since the stem is no longer rigidly connected to bone, neither by cementation nor by osseointegration, there is relative motion in axial direction of the stem with respect to bone. Since the stem still interacts with surrounding bone through normal contact, it contributes to stiffness and stability of the implant most notably for loading conditions with transverse forces and bending at the knee joint.
	
	Primary stability refers to the relative position stability on short terms after surgery, for cementless implantation by means of a press-fit which is continuously replaced in the healing process by secondary stability through osseointegration.
	%\footnote{For a fully cemented fixation this distinction between primary and secondary stability does not apply.}. 
	To enable osseointegration the restriction of micromotion to small values is crucial. This context elucidates why in the present concept, where the stem is not rigidly connected to bone, primary stability in terms of confined micromotion need not be strictly enforced. 
	%In contrast, the relative motion of the smooth stem implant at low friction coefficient is part of the concept. 
	
	%Ries et al. (2013): 
	%Aseptic loosening of the tibial component remains a major cause of failure in TKA and may be related, directly or indirectly, to micromotion [26]. Micromotion at the implant-cement or bone-cement interface causes generation of wear particles, which is supposed to be the main reason for aseptic loosening in the mid- and long term [11]. In the short-term, mechanical failure of implant fixation might be an overloaded bonecement or more likely a cement-implant failure. Therefore, the primary fixation marks one cornerstone for longevity of the prosthesis.
	
	For standard implantation, micromotion is frequently an indicator or even a major cause of aseptic loosening of the tibial component thus leading to TKA failure \cite{Vanlommel.2011}. But this is true for full cementation, where micromotion is not intended and induces at rough implant-cement-bone interfaces harmful wear particles \cite{Jacobs.1994}. This is in stark contrast to the present concept of a smooth stem enabling debris-free micromotion based on judiciously selected surface material as described in Sec.~\ref{subsec:realization}. 
	In this context it should be noted that the lower part of short stem hip endoprosthesis are typically smooth; they have the function to stabilize the position of the implant though they have no fixation to bone by osseointegration. Notice that suchlike design is observed for a multitude of implant models where the high primary stability is further enhanced by the rounded tip of the stem guided along the dorso-lateral cortex.
} 

%Rather than talking about primary stability why don't we talk about micro motion? One reason is that micro-motion is much more difficult to quantify. ... If we can limit the microscopic movement between implant and bone, hypothetically less than 100 $\mu$m (Brunski)  or  tested at 29 $\mu$m (Cameron, Pilliar), the osseointegration process may continue to occur. This whole concept of primary stability is a relative concept there is always instability it's just a question of how unstable unstable is?

%can be avoided by an interface layer between stem and bone similar to the cemented fixation, where a soft material is filled in which exhibits a low shear modulus and by virtue of incompressibility (Poisson contraction close to 0.5) a relatively high compression modulus. 

\subsection{Transferability and limitations} 
{\color{black}
	Since the plate-based force transfer mimics on purpose the pre-surgery  physiological conditions, it is justified to assume that the favorable characteristics are preserved no matter how patient-specific bone data or loading conditions may differ from the particular settings of the present work. Reliable conclusions clearly require further simulation results.  
	
	Remodeling depends on mechanical stimulus, biochemical processes, on age, sex and individual disposition showing considerable scatter between individuals. As a consequence, models for remodeling necessarily contain uncertainties and simulations cannot be predictive in a deterministic way. 
	
	The proposed fixation concept crucially relies on sufficient stiffness and strength of tibial bone {\color{black} in the resection plane, which is, however, not stronger loaded than in presurgery, physiological conditions.} In conclusion, the key premise to apply the proposed concept of avoiding stress-shielding along with bone resorption, is that other medical indications do not overrule this aspect.
	
	There are several simplifications in the model in terms of one single loading instead of a full gait analysis \cite{Galloway.2013}, \cite{Bergmann.2014}, thereby restricting to pure axial forces and e.g. no shear forces, no consideration of muscles and ligaments. Moreover, the cement mantle is assumed to exhibit constant thickness, clearly an idealization \cite{Taylor.2015}. The same is true for the assumption of sticking friction condition everywhere on the stem surface.  
	
	All findings are predictions from the lab in silico, since experimental results of the lab in vitro or even from clinical research in vivo are missing.   
}

\subsection{Implant fixation at knee compared to hip -- a broader perspective} 
The present work adopts the idea of \cite{Eidel.2018} for a novel fixation concept of short-stem hip implants. Therein, the concept is referred to as ''collar-cortex compression concept (CO$^4$)'' for describing the key mechanisms of force transmission for avoiding SED-reduction. The crucial aspect of modified interface conditions is transferable beyond the existing considerable differences between fixation of the plate-stem system into femur and tibia in view of different geometries, loading conditions, stiffness distributions to name but a few differences. For the hip implant however, the current generation of models typically do not exhibit a plate device while it is standard in TKA. Remarkably, the idea of the mechanical activation of cortical bone in the resection plane by a pre-stressed thrust-plate for a hip implant was introduced already by \cite{Huggler.1995}.    
 
\section{Conclusions} 

The main results and conclusions of the present work can be summarized as follows:

\begin{itemize}
	
	\item The stiffer the stem-bone interface, (i) the larger (smaller) the force fraction transmitted through the stem (plate), and (ii) the more pronounced the post-surgery SED reduction, hence stress-shielding. 
	
	\item Vice versa, force transmission primarily mediated through the tibial plate mechanically activates proximal tibia and confines SED reduction to an unprecedented extent both in intensity and spatially distal extension.
	
	\item Stress shielding can be avoided even for a metallic implant despite its large implant-to-bone stiffness mismatch, if the stem-bone interface is sufficiently compliant in terms of sliding friction conditions.
	
	\item The newly introduced force decomposition ratio of plate-stem is a reliable indicator of post-surgery SED changes. 
	
	\item The base-plate force transmission concept follows a bionics perspective in that 
	it mimics pre-surgery conditions of axial force transmission.
	
	\item For surgery practice in TKA the established standard set-up can be preserved; the implant geometry and materials, the cemented base-plate to bone interface and therefore all techniques and procedures. {\color{black} For revision surgery the novel concept of tibial implant fixation is favorable for its missing rigid connection of stem to bone.} 
	
	\item In spite the considerable differences in geometry, stiffness and loading conditions, the main characteristics of a recently proposed concept for the fixation of short stem hip endoprostheses in \cite{Eidel.2018} are transferable to the tibial implant fixation. 
	
\end{itemize}

\bigskip

{\bf Declarations of interest:} None. 

\bigskip

{\bf Funding:} This research did not receive any specific grant from funding agencies in the public, commercial, or not-for-profit sectors.

\bigskip
 
{\bf Acknowledgments:} BE acknowledges support by the Deutsche Forschungsgemeinschaft (DFG) within the Heisenberg program (grant no. EI 453/5-1).  

%=========================================================================================
 
\bibliography{TKA-bib-new}

% ------- layout-datei --------------
\bibliographystyle{plainnat}
%\bibliographystyle{plaindin}
%\bibliographystyle{plaindin_shortname2}
%\bibliographystyle{elsarticle-num-names}
% ------- bib-datei --------------
%\bibliography{js_master_2009}
%\begin{appendix}
%\include{appendixA}
%\end{appendix}
\end{document}